\documentclass[structabstract]{aa}   
\usepackage{graphicx}
\usepackage{txfonts}
\usepackage{natbib}
\bibpunct{(}{)}{;}{a}{}{,}

\begin{document}

\defcitealias{nz08}{NZ08}
\defcitealias{mf04}{MF04}

\title{GR models of the X-ray spectral variability of MCG--6-30-15}

\author{A. Nied\'zwiecki \inst{1} \and T. Miyakawa\inst{2}\inst{3}
   }

\institute{University of \L \'od\'z, Department of Astrophysics,
     Pomorska 149/153,  90-236 \L \'od\'z,
     Poland\\ 
     \email{niedzwiecki@uni.lodz.pl}        
     \and The Institute of Space and Astronautical Science,  Japan Aerospace
     Exploration Agency,3-1-1 Yoshinodai,  Sagamihara, Kanagawa
     229-8510, Japan 
     \and Department of Astronomy, Graduate School of
     Science,  University of Tokyo,7-3-1 Hongo, Bunkyo-ku, Tokyo
     113-0033, Japan }

\authorrunning{A. Nied\'zwiecki \& T. Miyakawa}

\titlerunning{GR models of the X-ray spectral variability of MCG--6-30-15}

  \date{Received 23 February 2009; accepted 09 October 2009}
 
\abstract 
{The extremely relativistic Fe lines, detected from some active galactic nuclei (AGN),
indicate that generation and reprocessing of the X-ray emission takes place 
in the immediate vicinity of the event horizon. Recently, general relativistic (GR) effects,
in particular light bending which is very strong in that region, have been  considered to
be the cause of  complex variability patterns observed in these AGNs.} 
{We study in detail the GR models of the X-ray spectral variability  
for various geometries of the X-ray source and with various relativistic effects 
being the dominant cause of spectral variability. The predicted properties
are compared with the observational data  of the Seyfert 1 galaxy MCG--6-30-15,
which is currently the best studied AGN with signatures of strong gravity effects.}  
{We focus on modeling the  root mean square (RMS) spectra. 
We compute the RMS spectra for the GR models using a Monte Carlo method,
and compare them with the RMS spectra from the
Suzaku observations of MCG--6-30-15 on January 2006.}  
{The data disfavor 
models with the X-ray source (1) moving vertically on 
the symmetry axis or (2) corotating with the disc and changing height 
not far above the disc surface. 
The most likely explanation for the observed fractional variability
is given by the model involving the X-ray source located
at a very small, varying distance from a rapidly 
rotating black hole. This model predicts some enhanced variations in the red wing
of the Fe line,
which are not seen in the Suzaku observations. However, the enhanced variability
of the red wing, while ruled out by the Suzaku data, is consistent
with an excess RMS variability, between 5 and 6 keV, reported for some
previous ASCA and XMM observations. We speculate that the presence or lack
of such a feature is related to the change of
the ionization state of the innermost part of the disc, however,
investigation of such effects is currently not possible
in our model (where a neutral disc is assumed). 
If the model, completed by a description of ionization effects,
proves to be fully consistent with the observational data,
it will provide a strong indication  that the central
black hole in MCG--6-30-15 rotates rapidly, supporting similar
conclusions derived from the Fe line profile.  } 
{}

\keywords{galaxies: individual: MCG--6-30-15 -- galaxies: active --
  galaxies: Seyfert -- X-rays: galaxies}

\maketitle

\label{firstpage}

\section{Introduction}
AGNs are powerful sources of strongly
variable X-ray emission, most likely  originating close to a central
black hole. The origin of the X-ray variability is poorly understood,
however, some of the observed variations are supposed to be directly
related to strong gravity effects and their investigation may give
insight into properties of the space-time metric in the vicinity of
the event horizon.

The  bright Seyfert 1 galaxy MCG--6-30-15 ($z=0.00775$) is the key object 
for such studies. Its X-ray spectrum shows an extremely distorted Fe K$\alpha$ line,
indicating that X-ray reprocessing takes place very
close to the central black hole \citep[e.g.][]{f02,m07}.
Furthermore, crucially for investigation of strong gravity in MCG--6-30-15,
radiation reflected from the inner disc is not
contaminated by  radiation reflected from distant matter (see Sect.\
\ref{sect:4.2}). The disc-line model for the detected line
profile requires that  roughly half of the observed Fe photons come
from the innermost parts of the accretion disc, within 6 gravitational
radii (these photons form the red wing of the line, below 6 keV) and
the remaining from a slightly more distant region, at a few tens of
gravitational radii (forming the blue peak, observed between 6 and 7
keV). 

The time-resolved X-ray spectra in MCG--6-30-15 indicate that the best
explanation of its spectral variability pattern  is given by a
phenomenological, two-component model  consisting of  (i) a highly
variable power-law continuum (referred to as primary emission) and
(ii) a much less variable,  and uncorrelated, component with a spectrum
characteristic of radiation reflected from the inner  disc
(e.g. \citealt{f03}, \citealt{vf04}, \citealt{l07};
an alternative explanation, proposed by \citealt{m08},
is briefly discussed
in Sect.\ \ref{sect:5.6}).  The latter component includes both the Fe K$\alpha$ line
and the associated Compton reflection hump, peaking around 30 keV,
produced by down-scattering of higher energy photons.  The varying
relative contributions of these spectral components can explain the
observed hardening  of the X-ray spectrum at lower fluxes as well as
reduced variations of the blue peak of the Fe K$\alpha$ line. 

The apparent lack of connection between the two components indicates
that some complex  physical mechanism should be involved, as the components are
supposed to originate together in the central region and a
strict correlation between them would be expected  in  a simple
reflection picture. \citet{f03} 
first argued that this
disconnectedness  may be explained by relativistic effects, in
particular by light bending and focusing of the primary emission
towards the accretion disc.   In this scenario, elaborated under some
specific assumptions by \citet[MF04]{mf04},
the change of the height of the primary X-ray source is the cause of spectral
variations.

However, a systematic investigation of this class of models  by
\citet[NZ08]{nz08}
questioned their ability to explain the observed effects. In particular,
MF04 assess the reduced variability of the
reflected component based on the behavior of the total flux of
fluorescent photons received by a distant observer. \citetalias{nz08}
point out that changes of the height of the source induce substantial changes in
the line profile,  with significant variations of the fluxes in the
blue peak and in the red wing.  These balance each other for some
range of parameters, reducing variations of the total flux,  however,
this property is not sufficient to explain the observed data. Then,
the model fails to reproduce the key property which motivated its
development, i.e. reduced variation of the blue peak of the Fe line. 

Interestingly,  \citetalias{nz08}
find an alternate model,  which can explain
reduced variations of the blue peak.  The model proposed  in NZ08
follows the generic idea of \citet{f03},
involving a
scenario where the change of position of the primary source leads to large
changes of directly observed emission and smaller changes of the amount
received by the disc. However, the model assumes changes of the radial
distance rather than the height and its  properties  rely on a
qualitatively distinct effect, namely bending to the equatorial plane
of the Kerr space-time. The model proposed in \citetalias{nz08}
predicts, however,  some enhanced
variations of the red wing of the Fe line, which may challenge its
applicability.

At present, the time-resolved spectra for sufficiently short ($\sim  10$ ksec) time bins
have poor quality, which does not allow us to study such fine
details  of spectral variability.  Thus, less direct tools are used. In
particular, the RMS spectra  -- showing the fractional variability as a
function of energy \citep[see e.g.][]{e02,m03,v03b}
-- allow us to investigate spectral variability in a model-independent manner.
  
In MCG--6-30-15, the general trend is that the RMS spectrum decreases
with the increase of energy, moreover, it has a
pronounced, broad depression around 6.5 keV. Such a shape of the RMS
spectrum  has been confirmed by the observational data from {\it ASCA} \citep{mifi03},
{\it XMM} \citep{vf04},
{\it RXTE} \citep{m03}
and {\it Suzaku} \citep{m07}. \citet{vf04}
show that this form of the RMS
spectrum may be reproduced in a model  with a variable power-law  and
a constant  reflection. Again, results of \citetalias{nz08}
challenge the physical motivation
for such a phenomenological description in terms of GR
models, as these typically predict some changes of the spectral shape of
reflected component. Changes in the Fe line profile have been studied,  
including their impact  on the RMS variability, in \citetalias{nz08}.
However, that paper neglected the Compton reflected component and its results 
cannot be directly compared with the observed data.

In this paper,  we extend the model developed in \citetalias{nz08} by including
the Compton reflected component, and we analyze  its predictions for
the fractional variability amplitude.  Then, we compare the simulated
RMS spectra with the spectra derived from three {\it Suzaku}
observations of MCG--6-30-15 performed in January 2006.

\section{GR models of the X-ray spectral variability}

\subsection{The GR models}
\label{sect:2.1}

We consider an accretion disc, surrounding a Kerr black hole,
irradiated by  X-rays emitted from an isotropic point source
(hereafter referred to  as the source of primary emission).  The black
hole is characterised by its mass, $M$, and angular momentum, $J$. We
use the Boyer-Lindquist coordinate system $x^i =
(t,R,\theta,\phi)$ and the following dimensionless parameters 
\begin{equation}
r = {R \over R_{\rm g}},  ~~~a = {J \over c R_{\rm g} M}, ~~~\Omega =
{{\rm d} \phi \over {\rm d} \hat t},
\end{equation}
where  $R_{\rm g} = GM/c^2$ is the gravitational radius and $\hat t =
ct/R_{\rm g}$. 

We assume that a geometrically thin, neutral,  optically thick disc is
located in the equatorial plane of the Kerr geometry. For distances
greater than the radius of the marginally stable circular orbit,
$r_{\rm ms}$, we assume circular motion of matter  forming the disc,
with Keplerian angular velocity \citep{bpt},
\begin{equation}
\Omega_{\rm K}(r) = {1 \over a+r^{3/2}}.
\end{equation}
We take into account reflection from matter which free falls within
$r_{\rm ms}$, however, this effect is negligible for the large values of
$a$ considered in this paper.  We assume the outer radius of the disc
$r_{\rm out}=1000$. Negligibly small dips, related to
this finite outer extent, occur in the simulated RMS spectra around
6.4 keV. 

Inclination of the line of sight  to the rotation axis of the black
hole is given by $\theta_{\rm obs}$.   Our discussion of properties
predicted by GR models, in Sect.\ \ref{sect:3}, focuses on values of the
inclination angle   between $\theta_{\rm obs}=25 \degr$  and $45
\degr$, which is relevant to modeling   MCG--6-30-15. The
location of the primary source is described by its Boyer-Lindquist coordinates,
$r_{\rm s}$ and  $\theta_{\rm s}$, or by $h_{\rm s} \equiv r_{\rm s} \cos \theta_{\rm s}$
and $\rho_{\rm s} \equiv r_{\rm s} \sin \theta_{\rm s}$.
Note that the azimuthal position of the source is not relevant
to our results, as we consider only spectra averaged over a complete orbit
of the source, see Sect.\ \ref{sect:2.4}.

We neglect transversal or radial motion of the source.  We take into account
the azimuthal motion with the generic assumption that the X-ray source (if displaced
from the symmetry axis) corotates with the disc.  We note, however, that the
description of such a corotating source is ambiguous at larger $h_{\rm s}$
and the choice of a specific assumption may significantly affect properties
of the model. Plausible  assumptions include relating the angular velocity of the source, 
$\Omega_{\rm s}(r_{\rm s},\theta_{\rm s})$, to the Keplerian angular velocity 
in the disc plane at $r_{\rm s}$ or at $\rho_{\rm s}$. 
In general, the former, with $\Omega_{\rm K}(r_{\rm s})$, yields an azimuthal velocity of the source,
$v_{\rm s}^{\phi}$, slightly smaller than the Keplerian velocity of the disc, 
$v_{\rm K}^{\phi}$ ($\simeq 0.5c$ in the innermost disc),
while the latter, with $\Omega_{\rm K}(\rho_{\rm s})$, yields $v_{\rm s}^{\phi}$ significantly 
larger than $v_{\rm K}^{\phi}$;
note that the azimuthal velocities, $v_{\rm s}^{\phi}$ and $v_{\rm K}^{\phi}$, are defined
with respect to the locally non-rotating frame, see \citet{bpt}.
In particular, for $h_{\rm s}/r_{\rm s} \simeq 1$, 
using $\Omega_{\rm s}=\Omega_{\rm K}(r_{\rm s})$ we get $v_{\rm s}^{\phi} \la 0.4c$, 
while $\Omega_{\rm s}=\Omega_{\rm K}(\rho_{\rm s})$ yields $v_{\rm s}^{\phi}$ exceeding
$0.6c$ and $0.8c$ at $r_{\rm s} = 3$ and $r_{\rm s} = 2$, respectively. 
The above issue is particularly relevant in model $C$, defined below, where
$\Omega_{\rm s}=\Omega_{\rm K}(\rho_{\rm s})$ is assumed (following \citetalias{mf04}).
In turn, our model $S$ (see below) 
assumes $\Omega_{\rm s}=\Omega_{\rm K}(r_{\rm s})$, although at small $h_{\rm s}/r_{\rm s}$, 
considered in this model, the difference between the two approaches is  negligible.

The relevant relativistic effects are clearly represented in the
following three models. 

{\it Model $S$} involves a source located close to the disc surface and rotating
with the Keplerian velocity, $\Omega_{\rm K}(r_{\rm s})$.  
 We consider very small heights of
the source above the disc surface, $h_{\rm s} \ll r_{\rm s}$, for which rigid coupling (and
corotation) of the source  is the most likely configuration; moreover, the description
of azimuthal motion is not subject to the ambiguities noted above. 
Variability
effects in model $S$ result from varying radial distance, $r_{\rm
  s}$, at a constant polar angle $\theta_{\rm s}$. Note that the change 
of the radial position leads to slight changes of the height, $h_{\rm s}$, so the source
motion is conical rather than plane parallel.  

Most results presented for model $S$ 
in this paper are derived for a fixed polar angle, $\theta_{\rm s}=1.5$ rad, yielding 
$h_{\rm s}/r_{\rm s}=0.07$; in Sect.\ \ref{sect:3.3} we show results for other
values of $\theta_{\rm s}$.

Two different physical effects can be studied in the following two regimes of model $S$.
At large distances, $r_{\rm s}>6$, the bulk of the reflected radiation
arises from a small spot  under the source; moreover, the gravitational redshift
is weak and the variability properties are dominated 
by local Doppler distortions, see Sect.\ \ref{sect:3.2}. For $r_{\rm s}<4$ and high
values of $a$ ($>0.9$), a qualitatively distinct
property of the Kerr metric (namely the bending of photon
trajectories to the equatorial plane) results  in both strong
and approximately constant illumination of the regions of the disc
where the blue  peak of the Fe line is formed, see \citetalias{nz08}.
The latter case, i.e. model $S$ with 
large $a$ and small $r_{\rm s}$, will be referred to as model $S^{\rm NZ}$.

{\it Model $A$} involves a static primary source located on the
symmetry axis. Such a scenario, often referred to as  a lamp-post
model, has been  considered in a number of papers, e.g.\ \citet{mm96}, \citet{ph97}. 
Variability in model $A$ is
induced by the changes of the source height, $h_{\rm s}$, above the
disc.  This  model allows us to investigate properties related to bending
of photon trajectories toward the center, a dominant effect for such a
location of the source.

Our {\it model $C$} corresponds to a variability
model with a cylindrical-like motion of the primary source, developed
by \citetalias{mf04} (note that their original model assumed a small
value of $r_{\rm out}=100$, which influenced some of their results, see
discussion in \citetalias{nz08}).  The model (often referred to as the
light-bending model) involves a source changing its height above the
disc, similarly to the lamp-post model, but displaced from the
axis, with the constant projected radial distance $\rho_{\rm s}$, 
and rotating around it.  Furthermore,
this model assumes that at each $h_{\rm s}$ the source has the same
angular velocity, 
$\Omega_{\rm K}(\rho_{\rm s})$. As noted above, the last assumption yields
$v_{\rm s}^{\phi}$ significantly exceeding the Keplerian velocity
of the disc. Moreover, the dependence of $v_{\rm s}^{\phi}$
on $h_{\rm s}$ is non-monotonic, which is crucial for the variability predicted by this
model, see \citetalias{nz08}.  Then, model $C$ allows us to study the impact of
kinematic effects, which are complementary  to the  gravity effects
underlying properties of models $A$ and $S$.  All results for model $C$
shown in this paper are derived with $\rho_{\rm s}=2$.

In models $A$ and $C$ we consider the maximum rotation of a black
hole, with $a=0.998$. For model $S^{\rm NZ}$ - which appears to be the
most feasible  to explain the observed data - we investigate
its dependence  on the value of $a$.

\subsection{Intrinsic luminosity}

\label{sect:2.2}

In models $A$ and $C$ we assume the same intrinsic luminosity of the
primary source at all $h_{\rm s}$. 
Model $S$, in Sect.\ \ref{sect:3.2}, also assumes an identical luminosity at
various $r_{\rm s}$.  For model $S^{\rm NZ}$ we focus, however, on a (physically
motivated) scenario
where the radial profile of  intrinsic luminosity, $L^{\rm PT}(r_{\rm s})$, 
follows the dissipation rate per unit area in a Keplerian disc \citep{pt74}.
All computations for models $A$ and $C$, and for model $S$ or $S^{\rm NZ}$ 
in Sects.\ \ref{sect:3.2}--\ref{sect:3.4}, are performed assuming
a constant (in time) luminosity at any given location.
For model $S^{\rm NZ}$, we consider two further modifications:

\noindent
(i) A power-law modulation of the radial luminosity profile, 
$r^{\beta}_{\rm s} \times L^{\rm PT}(r_{\rm s})$;

\noindent 
(ii) As a step toward more realistic modeling, we take into account variations (in time) of 
intrinsic  luminosity. In most cases we assume a Gaussian distribution of luminosities,
with the deviation $\sigma_{\rm L}$,  but we note that properties of the model depend 
on the assumed distribution function; details are given in Sect.\ \ref{sect:3.5}.

\subsection{Monte Carlo computation of the RMS spectra}  
\label{sect:2.3}

For each model, we compute the observed spectra for a number of
primary source positions, using the following grids of  $r_{\rm s}$ or $h_{\rm s}$.
In models $A$ and $C$ the heights are linearly spaced with $\Delta h_{\rm s} = 1$
at $h_{\rm s} \ge 3$ (so $h_{\rm s}=3$, 4, 5, ...) and 
$\Delta h_{\rm s} = 0.2$ at $h_{\rm s} < 3$.
In model $S$ the radial distances are linearly spaced with $\Delta r_{\rm s} = 1$;
however, in the regime of model 
$S^{\rm NZ}$ (i.e. at $r_{\rm s} \le 4$) we use $\Delta r_{\rm s} = 0.1$.

For a given position of the primary source, we use a Monte Carlo
method, involving a fully general relativistic treatment of photon
transfer in the Kerr space-time, to find spectra, observed by a
distant observer, of both the primary emission and the reflected
radiation.  A large number of photons are generated from the primary
source (for each position we typically trace $10^9$ photons)
with isotropic distribution of initial directions in the source
rest frame and a power-law distribution of photon energies,
characterised  by a photon spectral  index, $\Gamma$.  Results
presented in this paper correspond to $\Gamma=2$.  We have not found
any noticeable difference in the RMS spectra computed for various
spectral indices in the range relevant to MCG--6-30-15,
$\Gamma=2-2.3$ (indicated by modeling of time-resolved spectra,
e.g.\ \citealt{l07}).

For each photon, equations of motion are solved to find whether the
photon crosses the event horizon, hits the disc surface or escapes
directly to the distant observer.  For photons hitting the disc, we perform
a Monte Carlo  simulation of Compton reflection. Photons transferring
in the disc are subject to consecutive Compton scattering events
competing with absorption. We use  abundances of \citet{ag89}.  
For an isotropic illumination of the disc surface, our code
reproduces, in the disc rest frame, the reflection spectrum described
by the {\it pexrav} model \citep{mz95}.
Our simulation of Compton reflection allows us to take into account an
incidence-angle dependent irradiation of the disc.

For a photon hitting  the disc surface with energy $ > 7.1$ keV, we
generate  an iron K$\alpha$ photon, with energy 6.4 keV, emerging from
the disc.   The relative weight of the Fe photon is related  to the
initial energy and direction of the incident photon by the
quasi-analytic formula [eqs.~(4-6)] from \citet{gf91}.
Similarly to \citetalias{nz08}, we modify the original formula by
multiplying it by a factor of 1.3 to account for  elemental
abundances, consistent with those assumed for Compton reflection.  We
assume a limb-darkened emission of Fe K$\alpha$ photons, with
intensity $I(\mu_{\rm em}) \propto 1 + 2.06 \mu_{\rm em}$, where
$\mu_{\rm em} \equiv \cos \theta_{\rm em}$ and $\theta_{\rm em}$ is
the polar emission angle in the disc rest frame.  
As shown in \citetalias{nz08},
for a limb-brightened emission, only small, systematic changes -
related to the strength of the blue peak (see section 3.7 in \citetalias{nz08}) -
occur in the RMS spectra.  We solve
equations of motion in the Kerr metric for both the Fe K$\alpha$ and
Compton reflected photons.  We take into account reflection of photons
that return to the disc,  following the previous reflection.

We compute the observed spectra for a number, $N$,  of various
positions of the primary source and then we  determine the RMS
spectrum according to the following definition
\begin{equation}
{\rm RMS}(E) =  { 1\over <f(E)>} {\sqrt
  {\sum_{i=1}^{N}{[f_i(E)-<f(E)>]^2 \over N-1}} },
\label{rms}
\end{equation}
where  $f_i(E)$ is the  photon flux in the energy band, $E$,
corresponding to $i$-th position of the source and  $<f(E)>$ is the
average photon flux in this band for all $N$ positions.

In our basic analysis, in Sect.\ \ref{sect:3.1}--\ref{sect:3.3}, we assume that each spectrum, 
$f_i$, corresponds to a specific position of a point source and the RMS spectra
are derived taking into account all positions within a range specified
by its lower and upper ends, $r_{\rm min}$ and $r_{\rm max}$ or $h_{\rm min}$ and 
$h_{\rm max}$, respectively; this typically involves summation over $\sim 10$ positions 
of the primary source. 

In such a case (i.e. for spectra, $f_i$, from a single position), the impact of
relativistic effects on spectral variability is maximized (see Sect.\ \ref{sect:3.4}).
However, such a large magnitude of relativistic effects can be observationally 
revealed only under very specific conditions. Namely, the light curve should be sampled in
time-bins with a size  comparable to (or shorter than) the
time-scale of the change of position of a   primary source.   
Furthermore, approximation of  the hard X-ray emitting region by a single, point-like source
may be inadequate. Particularly in model $S$, multiple flares may occur simultaneously at
various sites on the disc. 

To estimate predictions of GR models for
more complex arrangements of the primary source, or larger time-bin sizes,
we study the RMS spectra for the model with the energy spectra, $f_i$, formed 
by contributions from several random positions of the source. 
We make a simplifying assumption that each basic spectrum, $f_i$, contains a contribution 
from the same  number, $n_X$, of locations.
Each location is randomly generated with a continuous probability distribution; then,
$h_{\rm s}$ or $r_{\rm s}$ from the grid of a given model, nearest to the generated value, 
is taken into account. In models $A$ and $C$, $h_{\rm s}$ is generated with a uniform probability 
distribution between $h_{\rm min}$ and $h_{\rm max}$. In model $S^{\rm NZ}$ we use 
the  probability density in the form of a power-law, 
$P (r_{\rm s}) \propto r_{\rm s}^{\delta}$. The RMS spectra in
these more complex cases are computed with $N=30$.

Summarizing the above, model $S^{\rm NZ}$, which is studied most thoroughly here, follows the kinematic assumptions 
of model $S$ and, unless otherwise specified, assumes a constant intrinsic luminosity
given by $L^{\rm PT}(r_{\rm s})$. Modifications of these basic assumptions are parametrized by:
$\beta$ - for the change of the radial luminosity profile; $\sigma_{\rm L}$ - for
variations of intrinsic luminosity at (each given) $r_{\rm s}$; and $\delta$ - for the radial
probability distribution.

\subsection{Azimuthal averaging; time delays}

\label{sect:2.4}

We ignore time delays between the primary and reprocessed
radiation, related  to light-travel times between the source and
various parts of the disc; moreover, we determine the RMS spectra
using the spectra, $f_i$, averaged over azimuthal angle.  We emphasize
that both of these issues are not important to  our analysis, which
focuses on effects observed in MCG--6-30-15 (harbouring a black hole
with a mass,  $M \simeq 3 \times  10^6 M_{\sun}$; \citealt{mch05})
on a time-scale of 10--100 ksec.

Regarding the time delays, we note that - in most cases -
almost all the reprocessed radiation originates from $r < 30$ and the
delays are smaller than 1 ksec.  Only  in model $C$ with $h_{\rm
  s}>10$, irradiation of the disc at $r \ga 100$ cannot be neglected,
for which time delays of a few ksec occur.   These would be
relevant for modeling some effects studied for short time-scales,
e.g. in point-to-point RMS spectra with  1 ksec bins \citep[cf.][]{vf04}. 
The time bins used to derive the RMS spectra in this
paper are an order of magnitude larger and any variations on the
time-scale of a few ksec would be grossly undersampled.

Similarly, in most cases, the orbital period of the source does not
exceed 1 ksec, which justifies  the azimuthally-averaged treatment.
The only exception involves  model $S$ with  large $r_{\rm s}$ ($\ga
10$). We neglect the detailed study of this regime, as it appears not
relevant in modeling the observed data. 

   \begin{figure}
   \centering \includegraphics[width=8cm]{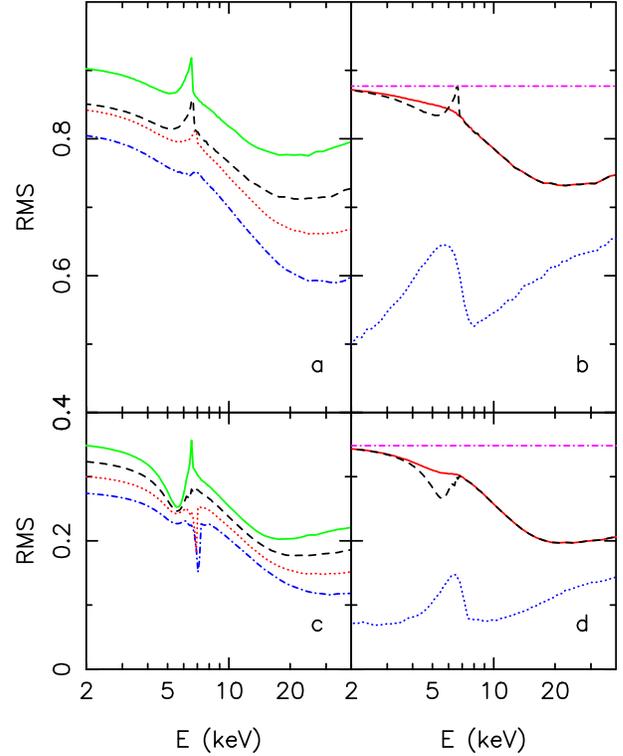}
      \caption{Panel (a) shows the RMS spectra for model $A$ with
        $h_{\rm s}$ changing between 1.8 and 10. Panel (c) shows the RMS
        spectra for model $C$  with $h_{\rm s}$  changing between 4
        and 20. In both panels, the spectra from top to bottom are for
        $\theta_{\rm obs}=25 \degr$ (solid, green online), $30 \degr$ (dashed, black; shifted
        vertically by -0.02, for clarity), $40 \degr$ (dotted, red; shifted
        vertically by -0.04) and $45  \degr$ (dot-dashed, blue; shifted vertically by
        -0.06).  Panels (b) and (d) show the RMS spectra for various
        spectral components in model A and C, respectively, with
        $\theta_{\rm obs}=30 \degr$ and other parameters indicated for
        panels (a) and (c). The curves, from bottom to top, are  for
        Compton reflected hump (dotted, blue; note that it does not include the Fe line),  primary +
        Compton reflection (solid, red) and  primary (power-law) component
        (dot-dashed, magenta). The dashed (black) curve is for the total emission,
        i.e. primary continuum + Compton reflection + Fe line.}
         \label{fig:f1}
   \end{figure}

   \begin{figure}
   \centering \includegraphics[width=6.5cm]{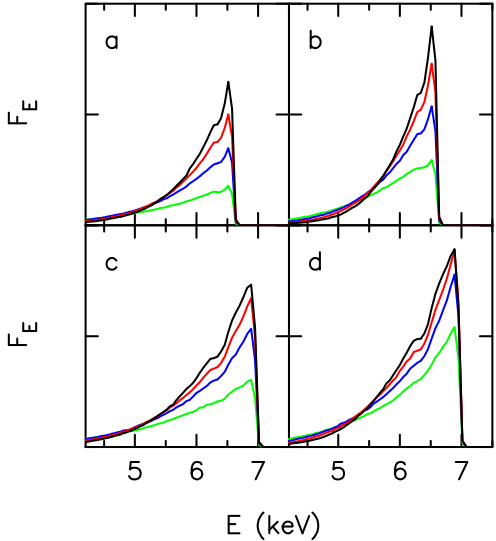}
      \caption{Changes of the Fe line profile in model $A$ (panels a and c) and model $C$
        (panels b and d). The top panels are  for $\theta_{\rm obs}=25 \degr$; 
        the bottom panels are  for $\theta_{\rm obs}=40 \degr$.
        In all panels, the profiles from bottom to top are for
        $h_{\rm s}=4$, 6, 8 and 10.   Note that, for $\theta_{\rm
          obs}=40 \degr$, rather subtle differences
        around the maximum of the blue peaks lead to opposite
        properties, i.e. a sharp excess or a sharp drop, around 6.9
        keV in the RMS spectra for model $A$ and $C$, respectively,
        see Fig.\ \ref{fig:f1}. The units are arbitrary but the
        same in all panels; profiles in panels (a) and (b) are rescaled by a factor of 0.55.}
         \label{fig:f2}
   \end{figure}

  \begin{figure}
   \centering \includegraphics[width=4.3cm]{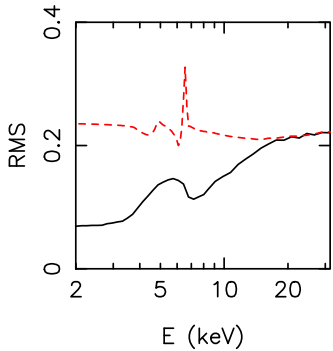}
      \caption{RMS spectra in two regimes where the fractional
        variability does not decline with increasing  energy. The
        bottom (solid, black online) is for model C with $h_{\rm s}$ changing
        between 0.2 and 1.2.  Only in this range of parameters is the
        reflected radiation more variable than the primary emission
        and a pronounced feature, related to the varying shape of the Fe
        edge, occurs in the RMS spectrum.  The upper (dashed, red) is for
        model S with $r_{\rm s}$ changing between 6 and 12.  Both
        spectra are for $\theta_{\rm obs}=30 \degr$.   }
         \label{fig:f3}
   \end{figure}

\section{Results}

\label{sect:3}

Figures \ref{fig:f1}, \ref{fig:f3}, \ref{fig:f4} and \ref{fig:f5} show
the RMS spectra for models $A$, $C$ and $S$, derived under the
assumption that each of the individual spectra, $f_i(E)$, contributing to the RMS
spectrum corresponds to a fixed, single position of a
primary (point) source. The RMS spectra are computed for
a constant intrinsic luminosity of the primary source.  Figures
\ref{fig:f6} and \ref{fig:f6new} show the RMS spectra for more complex scenarios,
discussed  in Sects.\ \ref{sect:3.4} and \ref{sect:3.5}.

\subsection{Changes of $h_{\rm s}$}
\label{sect:3.1}

As shown in Figs.\ \ref{fig:f1}(a)(c) and \ref{fig:f3}, models
assuming a vertical motion of the primary source do not predict a
depression around 6.5 keV, similar to that revealed in MCG--6-30-15 (cf.\ Fig.\ \ref{fig:f7} below).
Moreover, an opposite property, i.e. increase of fractional
variability at this energy, is typically predicted by these models in
most of the parameter space. Before discussing the specific models, 
we comment on the generic impact of various spectral components on
formation of the fractional variability spectrum, illustrated in
Fig.\ \ref{fig:f1}(b)(d). 
The narrow excesses in the RMS spectra
result from changes  of the Fe line profile. These are investigated 
in detail in NZ08; here we briefly discuss some basic properties
crucial for this study. We point out that relativistic distortions
of the Compton hump  are not as crucial for the fractional
variability  as distortions of discrete  spectral features.
This property results from a continuous energy distribution of the
hump and, in most cases, leads to decrease of the RMS spectrum, 
consistent with observations.

The primary continuum is
the most variable component, regarding variations of  the directly
observed flux,  but it does not change the spectral shape. Then, its
variability is represented by a flat RMS spectrum with a large
amplitude. In turn, changes of the amount of primary radiation
illuminating the disc  - resulting from the change of $h_{\rm s}$ -
are smaller (as a result of light bending). However, the distance  of
the most intensely illuminated area of the disc  changes with $h_{\rm
  s}$. As a result,  discrete spectral features in the reflected
radiation, in particular the Fe K$\alpha$ line and edge, are subject to
varying  relativistic distortions, yielding an excess fractional
variability around their rest energies. An excess, related to the 
variability  of the blue peak of the Fe K$\alpha$
line, is seen in most of the spectra in Fig.\ \ref{fig:f1}(a)(c)
(except for model $C$ with large $\theta_{\rm obs}$, see below).  As
we noted before, the excess variability caused by the Fe K$\alpha$
line occurs even when  the total flux in the line does not change
significantly.
 
Similarly, changes of the shape of  the Fe K$\alpha$ edge  result in
a pronounced excess, around 7 keV,  in the RMS spectra for the Compton
reflected component, see the bottom curves in
Fig.\  \ref{fig:f1}(b)(d).  However, variability at this energy is
dominated by the primary continuum and the Fe line, and  the excess
(due to the Fe edge) does not yield  any significant
signature in the RMS spectrum for the total observed radiation (except
for model $C$ with small $h_{\rm s}$, see below).  Crucially, though,
for interpretation of the observed data, the distortions of the Fe
edge yield a smooth RMS spectrum, see the middle solid curves,
for the sum of  the primary continuum and the Compton reflected radiation, 
in Fig.\ \ref{fig:f1}(b)(d), unlike the case of
reflection  from a static slab, for which a sharp drop occurs at 7.1
keV (cf Fig.\ \ref{fig:f7} below). 

The Compton hump, although subject to the same relativistic
distortions as the Fe line, is less significantly affected by their changing amount,
owing to the continuous spectral distribution of the hump. As a result, the amplitude of
the RMS spectrum for the Compton reflected component is typically much
smaller than  for the primary emission, compare the top and bottom
curves in Figs.\  \ref{fig:f1}(b)(d).  As the contribution of the
Compton reflected radiation to the total spectrum increases between 2 and 30 keV, 
the RMS spectrum for the total emission decreases, reaching a minimum around 30
keV where the reflection hump peaks.

\subsubsection{Model $A$: light-bending (to the center)}
\label{sect:3.1.1}

Trajectories of photons emitted from a source located close to the
symmetry axis are subject to a simple bending to the center,
independent of the value of $a$ (contrary to bending to the 
equatorial plane, discussed in Sect.\ \ref{sect:3.3}).  Models involving a
vertically moving source close to the axis predict strong variations
of the observed flux. They result from changing amounts of purely GR effects, i.e. the
light bending and gravitational time delay. However, these models
robustly predict an enhanced  variability around the Fe K$\alpha$
energy, see Fig.\ \ref{fig:f1}(a). This property, inconsistent with observations, property
results from the fact that 
a source located at a larger $h_{\rm s}$ illuminates a more extended
area of the disc and, therefore, the increase of $h_{\rm s}$ gives rise to rather significant
strengthening of the blue peak (see Fig.\ \ref{fig:f2}a).  The
varying strength of the peak yields a pronounced excess in the RMS
spectrum between 6 and 7 keV.  

Dependence on $\theta_{\rm obs}$, seen in Fig.\ \ref{fig:f1}(a), results
from gravitational focusing and Doppler collimation, toward observers with high
inclination angles, of radiation reflected from the innermost parts of the disc
(which receive most of irradiating flux from a source located at
small $h_{\rm s}$). A stronger contribution of the Compton reflected radiation for
larger $\theta_{\rm obs}$ gives rise to a stronger decrease of the RMS spectra with
increasing energy. 
The focusing toward higher $\theta_{\rm obs}$ results also in weaker variations
of the blue peak (compare Figs.\ \ref{fig:f2}(a) and \ref{fig:f2}(b)) and, thus, 
in a less pronounced  RMS excess around 6.4 keV.

\subsubsection{Model $C$: azimuthal motion}
 \label{sect:3.1.2}

The major difference between models $A$ and $C$ results from rotation
of the primary source around the symmetry axis in the latter model.
Then, in model $C$, changes in the fluxes of the directly observed and the irradiating
radiation, induced by the change of $h_{\rm s}$, result from the combination 
of light bending and Doppler  beaming related to this motion (we recall that in both models 
$A$ and $C$ the intrinsic luminosity remains constant). The specific kinematic
assumptions of model $C$ yield a non-monotonic dependence of azimuthal
velocity on  $h_{\rm s}$, leading to different properties of the model
in two ranges of $h_{\rm s}$ (see \citetalias{nz08} for details). 
The critical height, approximately limiting these two ranges, is the one 
at which the azimuthal velocity achieves a maximum value (for $\rho_{\rm
  s}=2$ this maximum occurs at $h_{\rm s}=2$).
At small  $h_{\rm s}$ ($ < 2$ for $\rho_{\rm s}=2$),   
variations of the directly  observed primary continuum are
strongly reduced and the RMS spectrum is shaped by spectral changes of
the Compton hump; note that the reduced variations of the primary flux 
are observed only at smaller $\theta_{\rm obs}$ (i.e.\ these considered here) and result 
from more efficient Doppler beaming to larger $\theta_{\rm obs}$, due to the increase 
of $v_{\rm s}^{\phi}$, which approximately balances less efficient light bending 
with the increase of $h_{\rm s}$. The predicted increase of the RMS variability for
increasing energies, accompanied by excess variability around 6 keV
(caused by the changing Fe K$\alpha$ edge), see
Fig.\  \ref{fig:f3}, is clearly not consistent with observed data.

At larger heights, the trends in both components are reversed. The
primary continuum is now  the most variable component, yielding the
decrease of the RMS spectrum towards higher energies.  Effects related
to changes of the line profile strongly depend on $\theta_{\rm obs}$.
The crucial property resulting from the Doppler beaming of primary emission is that the
reflected radiation  originates from more extended regions of the disc
(with $r > h_{\rm s}$, as opposed to model $A$) and hence the Fe line
is somewhat less variable than in model $A$ (see
Fig.\ \ref{fig:f2}).  This is reflected in the RMS spectrum by the
sharp drop, with the width of a few hundred eV, occurring for $\theta_{\rm obs}>30 \degr$
 at the energy of the maximum of the blue peak.
Regardless of the value of $\theta_{\rm obs}$, model $C$ predicts a
pronounced depression around 5.5 keV.
 
   \begin{figure}
   \centering \includegraphics[width=8cm]{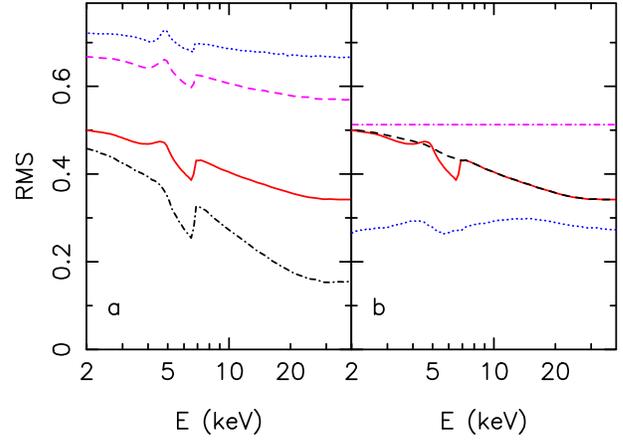}
      \caption{Panel (a) shows the RMS spectra for model $S^{\rm NZ}$
        with different values of spin parameter $a=0.9$ (dotted, blue online), 0.95
        (dashed, magenta), 0.98 (solid, red) and 0.998 (dot-dashed, black), from top to bottom.
        For all curves, $r_{\rm s}$ varies between $r_{\rm min} = r_{\rm ms}(a)$ and
        $r_{\rm max} =3.2$, and $\theta_{\rm obs}=30 \degr$.  Panel (b) shows the RMS
        spectra for various spectral components, as in
        Fig.\ \ref{fig:f1}(b)(d), in the model with $a=0.98$.  }
         \label{fig:f4}
   \end{figure}

  \begin{figure}
   \centering \includegraphics[width=8cm]{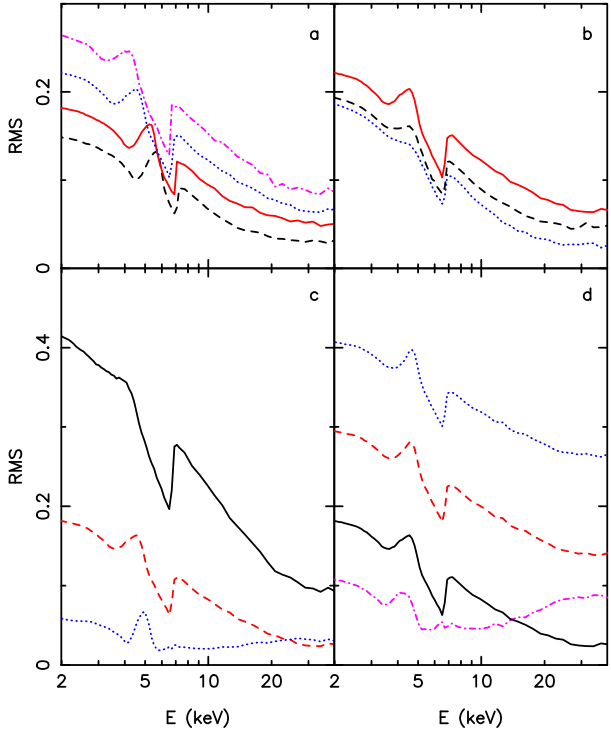}
      \caption{Dependence of the RMS spectra in model $S^{\rm NZ}$ on:
        $\theta_{\rm obs}$ (a); $\theta_{\rm s}$, or equivalently $h_{\rm s}/r_{\rm s}$ (b); 
        $r_{\rm min}$ and $r_{\rm max}$ (c); $\beta$ (d).
        All spectra are for $a= 0.998$; panels (b-d) are for $\theta_{\rm obs}=30 \degr$;
        panels (a),(b) and (d) are for $r_{\rm min}=2$ and $r_{\rm max}=3$.  
        Panel (a):  $\theta_{\rm obs}=25 \degr$ (dot-dashed, magenta online; shifted
        vertically by 0.06), $30 \degr$ (dotted, blue; shifted by 0.04), $40 \degr$
        (solid, red; shifted by 0.02) and $45 \degr$ (dashed, black).  
        Panel (b): the spectra from top
        to bottom are for $h_{\rm s}/ r_{\rm s}=0.07$ (solid, red; shifted
        vertically by 0.04), $h_{\rm s}/ r_{\rm s}=0.14$ (dashed, black; shifted
        vertically by 0.02) and  $h_{\rm s}/ r_{\rm s}=0.21$ (dotted, blue).
        Panel (c): the spectra from top to
        bottom are for $r_{\rm s}$ varying between $r_{\rm min}=1.6$ and $r_{\rm max}=2.6$
        (solid, black), $r_{\rm min}=2$ and $r_{\rm max}=3$ (dashed, red), 
        $r_{\rm min}=2.4$ and $r_{\rm max}=3.4$ (dotted, blue).  
        Panel (d): the spectra from top to
        bottom are for $\beta=2$ (dotted, blue), $\beta=1$ (dashed, red), $\beta=0$
        (solid, black), $\beta=-0.7$ (dot-dashed, magenta).}
   \label{fig:f5}
   \end{figure}

\subsection{Model $S$, large $r_{\rm s}$: local Doppler effects}

\label{sect:3.2}

The upper curve in Fig.\ \ref{fig:f3} shows the RMS spectrum for model
$S$ with large $r_{\rm s}$ ($\ge 6$). Radiation reprocessed from
the emission of a source located close to the disc surface at large
$r_{\rm s}$ originates in bulk from a small  spot below the source.
In such a case, both the directly observed primary emission and the
reflected radiation are subject to the same relativistic distortions
and, therefore, the relative normalization of the latter component is
the same  as for a static slab illuminated in flat space-time.  The
departure of the RMS spectrum from a flat form results from varying
Doppler distortion of the Fe line, corresponding to changes of $r_{\rm
  s}$.  These changes yield a strong excess around 6.4 keV due to the change
of the energy of the maximum of the blue peak. An additional, smaller excess  related
to the varying extent of the line occurs at lower energies.

Note, however, that such a strong excess corresponds only to a
simplified scenario with $n_X=1$ (see Sect.\ \ref{sect:3.4}).

\subsection{Model $S^{\rm NZ}$ (small $r_{\rm s}$ in the Kerr metric): bending to the equatorial plane}

\label{sect:3.3}

Figures \ref{fig:f4} and \ref{fig:f5} show the RMS spectra for model
$S^{\rm NZ}$ with small, {\it varying} $r_{\rm s}$ and large $a$.
Such parameters define a unique regime with the reduced variability of 
the blue peak of the  line occurring independently of other specific
assumptions.  In this scenario,  the reflected radiation has two
components with different variability behavior. The first one arises
locally from a strongly  irradiated spot (or - after azimuthal
averaging - a narrow ring) under the primary source. In principle,
this component follows the behavior  described in Sect.\ \ref{sect:3.2}, but -
in addition - it is subject to strong (and varying with $r_{\rm s}$)
gravitational redshift. The variable redshift of the Fe line gives
rise to a pronounced excess in the RMS spectrum between 4 and 6 keV.
The second component arises from a slightly more  distant region, at
$\sim 10 R_{\rm g}$, which is strongly illuminated due to bending to
the equatorial plane, an effect significantly affecting radiation
emitted from $r_{\rm s} < 4$ for $a > 0.9$. It is the second component
which yields the unique properties of this model, through the combination
of the following two effects: (1) the observed blue peak is formed
mostly by photons emitted at $\sim 10 R_{\rm g}$; and (2) the flux
illuminating that site  remains approximately constant while $r_{\rm
  s}$ changes (see NZ08 for details).  The resulting reduction of
variations of the blue peak yields a pronounced depression in the RMS
spectrum between 6 and 7 keV.  As the reduction effect fully relies on
the properties of the Kerr metric, it is obviously stronger for larger
$a$, see Fig.\ \ref{fig:f4}(a). 

Similarly to model $A$ and model $C$ with  $h_{\rm s} >
2$, the Compton component is less variable than the primary continuum, see
Fig.\ \ref{fig:f4}(b), and as a result the RMS spectrum decreases with
increasing energy.

The most significant change in the RMS spectrum, corresponding to the
change of $\theta_{\rm obs}$, is related to the shape and location of
the excess between 4 and 6 keV (produced by variations of the red wing), see
Fig.\ \ref{fig:f5}(a). Similarly, the change of $\theta_{\rm s}$  
is most significantly reflected in the shape of the
excess. For smaller $\theta_{\rm s}$, i.e.\ larger $h_{\rm s}/r_{\rm s}$,
the hot spot under the source, where the variable part of the red wing
is formed, is more extended and therefore the excess is less
pronounced, see Fig.\ \ref{fig:f5}(b). 

In model $S^{\rm NZ}$, the RMS spectrum is extremely sensitive to even small 
changes in $r_{\rm min}$ and $r_{\rm max}$, see
Fig.\ \ref{fig:f5}(c). For $r_{\rm min} \ge 2.4$, bending to the equatorial plane
is too weak to give rise to a significant reflection component from larger $r$ 
and the RMS is flat above 6 keV.   

Figure \ref{fig:f5}(d) illustrates changes of the RMS spectrum
resulting from the change of the radial profile of intrinsic luminosity,
with the power-law modification parametrized by $\beta$, defined in Sect.\ \ref{sect:2.2}.
Interestingly, less centrally concentrated profiles, with $\beta>0$,
preserve the shape of the RMS spectrum with only a slight change in the shape 
of the excess between 4 and 6;
however, the amplitude of the spectrum increases significantly with the increase of $\beta$.
For $\beta<0$, the RMS spectra flatten; furthermore, for $\beta < -0.5$ their qualitative properties 
change, see the dot-dashed curve. In particular, no drop around 6.4 keV occurs,
 and the spectrum increases at $E>10$ keV. Note that such negative values of $\beta$ 
would characterise discs with non-zero stress at the marginally stable orbit, considered e.g.\ by 
\citet{k99}.

   \begin{figure}
   \centering \includegraphics[width=8cm]{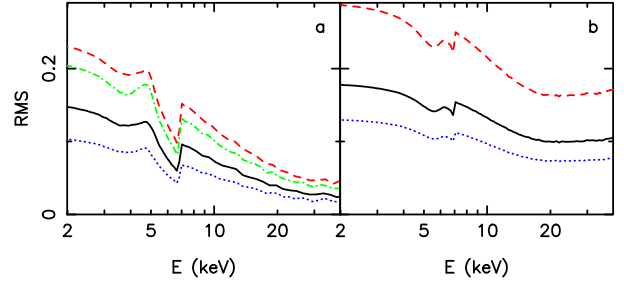}
      \caption{ Changes in the RMS spectrum
        corresponding to the increase of the number, $n_{\rm X}$, of the source
        positions contributing to the energy spectra.  
        Panel (a): model $S^{\rm NZ}$ with $\delta=0$, $r_{\rm min}=1.8$ and $r_{\rm max}=3$. 
        Panel (b): model $C$ with $h_{\rm min}=4$
        and $h_{\rm max}=20$. Models in both panels assume $a=0.998$ and
        $\theta_{\rm obs}=35 \degr$.
        In both panels, the dashed (red online), solid (black) and
        dotted (blue) curves  correspond to  $n_{\rm X}=1$, 3 and 5,
        respectively.    The dot-dashed (green) curve in panel (a) is for a single ($n_{\rm X}=1$)
        primary source with the size $\Delta r_{\rm s}=0.2$.}
         \label{fig:f6}
   \end{figure}

\subsection{More complex arrangements of the primary source}

\label{sect:3.4}

We consider here the RMS spectra constructed from the energy spectra, $f_i$, formed
by contributions from several positions of the source.   
A model with compact flares occurring simultaneously at various random
locations is an obvious example requiring such treatment.
An equivalent effect should occur, regardless of the configuration of
the emitting region, in any model involving changes of location,  
for the increase of the size of time bins.  

\begin{figure}
   \centering \includegraphics[width=9cm]{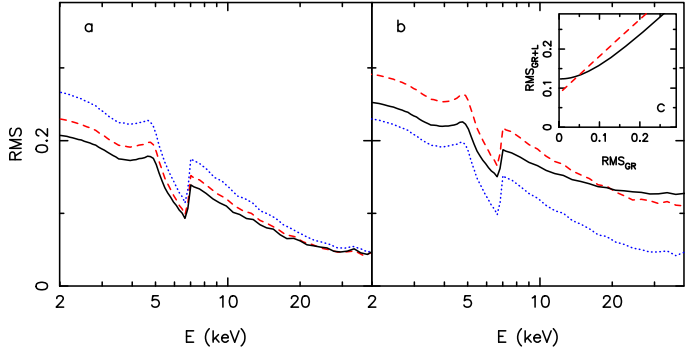}
      \caption{Panels (a) and (b) are for model
        $S^{\rm NZ}$ with $\theta_{\rm obs}=35 \degr$, $a=0.998$, $r_{\rm min}=1.8$, 
        $r_{\rm max}=3$ and $n_{\rm X}=1$. Panel (a): the spectra from top to bottom are 
        for $\delta=-1.5$ (dotted, blue online), $\delta=0$ (dashed, red) and $\delta=1.5$ 
        (solid, black). Panel (b): 
        The dotted (blue) curve is for constant intrinsic luminosity,
        $L^{\rm PT}(r_{\rm s})$, at given $r_{\rm s}$;  the solid
        (black) curve is for intrinsic luminosity generated from
        the Gaussian distribution with $\sigma_{\rm L}(r_{\rm s}) = 0.15 L^{\rm PT}(r_{\rm s})$;
        the dashed (red) curve is for intrinsic luminosity generated from
        a uniform distribution with $s = 0.15$.
        The inner panel (c) shows the RMS amplitude in a model involving 
        changes of intrinsic luminosity as a function of the RMS variability 
        without such changes; the dashed (red) and the solid (black) curves
        are for the uniform and Gaussian distribution, respectively, both 
        with $s = 0.15$; see text for details. }
      \label{fig:f6new}
   \end{figure}

The energy spectra in this (and the following) section are computed as a
mixture of spectra from randomly generated positions. In Fig.\ \ref{fig:f6} the 
positions are drawn with uniform probability and in Fig.\ \ref{fig:f6new}(a) 
we illustrate the effects of the change of the probability distribution.

Figures \ref{fig:f6}(a)(b) show changes in the RMS spectra resulting
from the increase of the number of source positions, $n_X$, contributing to an
individual spectrum, $f_i$. Obviously, deviation between the
average spectrum  and each of the individual spectra decreases, with the increase of $n_X$, which reduces
both the RMS amplitude and the dependence of the RMS spectrum on energy.

An essentially similar, but less pronounced, effect occurs for an increase of the size, $\Delta r_{\rm
  s}$, of the primary source, see the dot-dashed curve in Fig.\ \ref{fig:f6}(a); emission of an
extended source is approximated by superposition  of emissions from
point sources located in the range [$r_{\rm s} - 0.5 \Delta r_{\rm
  s}$, $r_{\rm s} + 0.5 \Delta r_{\rm s}$], so the energy spectra for
$\Delta r_{\rm s}=0.2$ are formed by adding spectra from 3 adjacent positions of
the model $S^{\rm NZ}$ grid.  Again, deviations
between spectra from more extended primary sources are smaller than in
models with a single point source. However, differences between spectra are more systematic,
and larger, than between those resulting from $n_{\rm X}>1$ with randomly generated $r_{\rm s}$
and, therefore, the flattening effect is weaker.

Note that the change of the radial probability density (which effectively modulates the 
radial emissivity profile) has a different effect on the RMS spectrum than the change of 
the radial luminosity profile. In particular, the
dominating contribution from smaller $r_{\rm s}$ (corresponding to $\delta<0$) 
does not result in a qualitative change 
of the RMS spectrum (contrary to the modification with $\beta < 0$), see  
Fig.\  \ref{fig:f6new}(a). 
    
Our model for the generation of active regions, with the same number of source
locations contributing to each spectrum, and an implicit assumption
of the same duration of emission at each position of the source, is certainly oversimplified;
see e.g.\ \citet{p08}
for a more sophisticated modeling 
of the generation of flare-like features in random processes.
Our main purpose here is to illustrate flattening of the RMS spectrum,
which is a generic trend corresponding to the increase of $n_X$.
See also \citet{cz04}
for the RMS spectra from their model of 
a spotted disc, which are much flatter than those made here with small
$n_X$ (however, their model neglects effects  of transfer from the X-ray source to the
disc, which is a qualitative difference to our model $S^{\rm NZ}$).

\subsection{Changes of intrinsic luminosity}

\label{sect:3.5}

Finally, we take into account changes (in time) of intrinsic luminosity.
The ability to describe the entire variability pattern in terms of relativistic effects
is an attractive property of GR models, however, assumption of a strictly constant luminosity
seems rather unrealistic. Results of the previous sections, though, could be
 applied directly to modeling a system with the time-scale of intrinsic variations  much shorter than
the time-bin length (so that each bin probes emission
averaged  over various luminosity states).

Figure \ref{fig:f6new}(b) shows the RMS spectra for
model $S^{\rm NZ}$, combining changes of the radial location with changes of
the intrinsic luminosity (at each given $r_{\rm s}$). The location $r_{\rm s}$ is randomly
generated (with  $\delta=0$) and for each $r_{\rm s}$ we generate a random
value of luminosity. Thus, the generated sequence of basic energy spectra typically contains several
contributions from a given location, each with a different normalization.   

It is interesting to notice that 
the results depend on the assumed distribution of luminosities.
In order to illustrate this, we construct the RMS spectrum using: 

\noindent
(i) a Gaussian 
distribution, $P(L) \propto \exp[-0.5(L-L_0)^2/\sigma_{\rm L}^2]$, where 
$L_0=L^{\rm PT}(r_{\rm s})$ and $\sigma_{\rm L}=s L^{\rm PT}(r_{\rm s})$;

\noindent
(ii) a uniform distribution in the range 
$[L_0(1-s),L_0(1+s)]$, with
$L_0=L^{\rm PT}(r_{\rm s})$.

\noindent
In both cases, $s$ is a constant number (independent of $r_{\rm s}$), thus, both 
the deviation for distribution (i) and the amplitude of variations for distribution (ii) 
change slightly with $r_{\rm s}$.  In general, changes of intrinsic luminosity
increase the amplitude of the RMS spectrum, however, the Gaussian distribution leads 
also to flattening, while the uniform distribution does not change the shape of this spectrum
with respect to that from the model with constant luminosity.

The relevant property underlying the above difference is illustrated in 
Fig.\ \ref{fig:f6new}(c). The figure
shows the relation between the RMS computed with (${\rm RMS}_{\rm GR+L}$) 
and without (${\rm RMS}_{\rm GR}$) changes of intrinsic luminosity for a simple model
using two random quantities, $x_F$ and $x_L$.
$x_F$ is generated using a uniform distribution and represents the received flux
in GR models with constant luminosity.
$x_L$  represents the intrinsic luminosity and is generated using a Gaussian (the solid curve) 
or uniform (the dashed curve) distribution, as defined above, both with $s=0.15$.
${\rm RMS}_{\rm GR}$ is constructed using the generated values of $x_F$,
the increase of the amplitude of variations of $x_F$ yields larger ${\rm RMS}_{\rm GR}$.
${\rm RMS}_{\rm GR+L}$ is constructed using the products of $x_F$ and $x_L$.
Note that the uniform distribution of $x_L$ leads to a linear relation between 
${\rm RMS}_{\rm GR+L}$ and ${\rm RMS}_{\rm GR}$, which is reflected in the unchanged shape of 
the spectrum in Fig.\ \ref{fig:f6new}(b), while for the  Gaussian distribution
this relation flattens at small ${\rm RMS}_{\rm GR}$.

\section{Application to MCG--6-30-15}

\label{sect:4}

\subsection{The Suzaku observations}

\label{sect:4.1}

Because of the wide bandpass coverage of energies provided by detectors
on-board  {\it Suzaku}, it is currently the most suitable X-ray
observatory for testing relativistic reflection models.  In
particular, crucial information comes from the {\it Hard X-ray
  Detector} data, allowing us to measure the Compton reflection
hump simultaneously with the Fe K$\alpha$ line.  MCG--6-30-15 was
observed three times by {\it Suzaku} in 2006 January  in a state
typical for this source, regarding the mean X-ray flux and variations
of the flux by a factor of $\sim 2$ in a few ksec \citep[see][]{m07}.
Crucially for our analysis, \citet{m07} 
find for these observations  that (1) the average profile of the Fe
K$\alpha$ line has a relativistic shape, with the pronounced blue
peak and the red wing extending down to $\simeq 4$ keV,  consistent
with that from the long {\it XMM} observation in 2001 \citep{f02};
(2) the average relative normalization  of the Compton
hump, given by the usual parameter $R$ ($R=1$ corresponds to the
strength of reflection  expected from a reflector subtending $2 \pi$
sr at an isotropic primary source), is $R \simeq 3.8$, consistent with
previous estimations from {\it Beppo-SAX} \citep{b03},
and rises to  $R \approx 5$ in the low flux spectra; (3)  on a
time-scale of tens of ksec, spectral changes are consistent with the
phenomenological, two-component model.

\begin{figure}
   \centering \includegraphics[width=7cm]{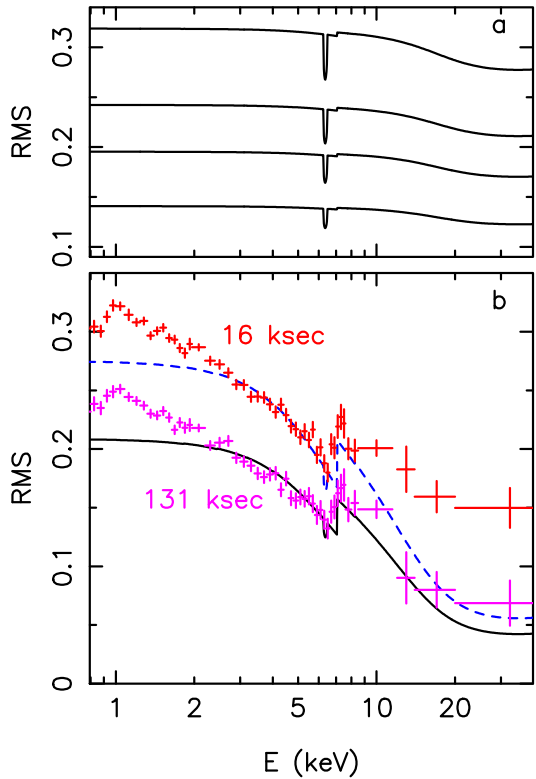}
      \caption{All curves show the RMS spectra for a model  with a
        constant Compton reflection (derived with {\it pexrav}), a
        constant line at 6.4 keV with EW = 30 eV (and with
        $\sigma=0.1$ keV, matching the width of the data energy bin)
        and a varying power-law  component.  Panel a: the RMS spectra
        are computed with various amplitudes of the power-law variability,
        but all have the same $R=0.15$,  approximately consistent with
        the EW of the line.  Panel b: the simulated RMS spectra, with
        $R=4$, compared with the {\it Suzaku} data.  The upper (red)
        and lower (magenta) points show the RMS spectra  for the {\it
          Suzaku} observations with a time bin size of 16 ksec and
        131 ksec, respectively. }
      \label{fig:f7}
   \end{figure}

We use the data from three observations of MCG--6-30-15 by the {\it
  Suzaku} satellite on 2006 January  9--14 (143 ksec exposure), 23--26
(99 ksec) and 27--30 (97 ksec).  The total on-orbit time is about 780
ksec.  For data reduction, we
have used the HEADAS 6.5 software package provided by NASA/GSFC.   We
determined  the RMS amplitude, and its error, as a function of
energy, in the standard manner (e.g.\ \citealt{v03b};
the definition is analogous to equation (\ref{rms}) but contains an
additional term related to the count rate error). In this paper
we use the RMS spectra for two time-bin sizes, approximately 16 ksec and
131 ksec (more specifically $2^{14}$ sec and $2^{17}$ sec),  
with approximately 7 and 50 ksec of
exposure time per bin, respectively.  The RMS spectra for 16 ksec
and 131 ksec are shown in Fig.\ \ref{fig:f7}(b) and all panels of
Fig.\ \ref{fig:f8} by the upper and lower set of points,
respectively.

The RMS spectra indicate strong decrease of variability at $< 1$ keV,
however, modeling of this band is beyond the scope of this paper.  The
decrease  most likely results from the presence of  a soft X-ray
excess, which is clearly seen in the average spectra of MCG--6-30-15
as well as in the spectrum of the non-varying spectral component (see
e.g. figs. 3 and 22 in \citealt{vf04}).
The origin of such soft excesses, typically observed in AGNs, is not clear 
\citep[see e.g.][]{s07}.
Remarkably, one of the possible explanations, involving the relativistically 
blurred photoionized disc reflection (where the soft excess is explained as 
being composed of many broad lines, see \citealt{c06}),
is consistent with the scenario discussed in Sect.\ \ref{sect:5.5}.
 
\subsection{Non-relativistic reflection}

\label{sect:4.2}

Before applying the GR models, we briefly comment on the contribution of
reflection from distant matter, which for some other objects is
considered as the explanation for RMS spectra similar to these derived
in MCG--6-30-15 \citep[e.g.][]{t09}.
In MCG--6-30-15,
however, such a component should have a marginal effect,  as indicated
by the very small contribution of a narrow Fe K$\alpha$ line at 6.4 keV
\citep[e.g.][]{i96,l02}. 

\begin{figure*}
   \centering \includegraphics[width=17cm]{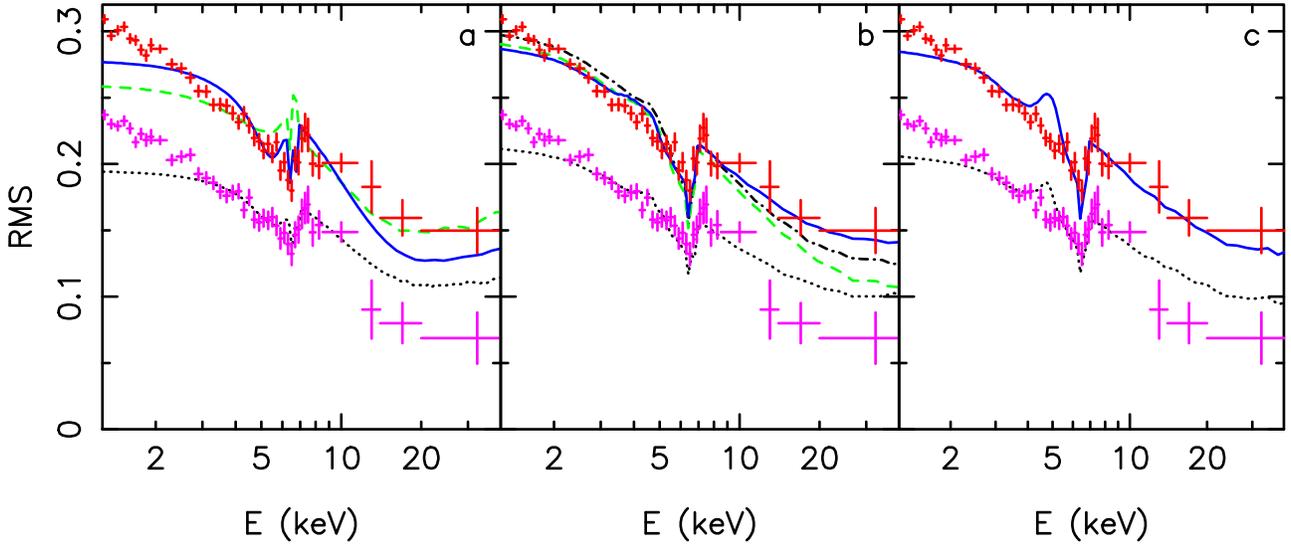}
      \caption{Fits of the {\it Suzaku} RMS spectra with the GR
        models.  In all panels, the upper (red online) and lower (magenta)
        points show the {\it Suzaku} RMS spectra  for   16 ksec and
        131 ksec, respectively. All models involve a constant line at 6.4 keV  with
        the ${\rm EW}=30$ eV and $\sigma=0.1$ keV; all models assume
        $a=0.998$ and $\theta_{\rm obs}=35 \degr$.   Panel (a):
        models involving vertical motion of the primary source.  The
        dashed (green online; shifted vertically by $-0.4$) curve is  
        for model $A$  with $h_{\rm min} = 2.2$ and 
        $h_{\rm max} = 10$. The solid (blue; shifted vertically by $-0.07$) and 
        dotted (black;  shifted vertically by $0.01$) curves are for model $C$, with
        $h_{\rm min} = 4$ and $h_{\rm max} = 20$, for $n_{\rm X}=1$ and 3,
        respectively. Panel (b): the solid (blue) and dotted (black) curves are for 
        $n_{\rm X}=3$ and 5, respectively, in model $S^{\rm NZ}$ with
        $r_{\rm min} = 1.6$, $r_{\rm max} = 3$, $\delta=-1.5$, $\sigma_{\rm L}=0.2L^{\rm PT}$
        (and $\beta=0$). The dot-dashed (black) curve is for the same parameters as the solid 
        curve, except for  (larger) $r_{\rm max} = 3.4$.
        The dashed (green) curve is for $r_{\rm min} = 1.4$ and $r_{\rm max} = 3$; 
        $n_{\rm X}=3$, $\beta=0$, $\delta=0$, $\sigma_g=0$.
        Panel (c): model $S^{\rm NZ}$ with $r_{\rm min} = 1.8$ and $r_{\rm max} = 3$, 
        $\beta=1$ ($\delta=0$ and $\sigma_{\rm L}=0$); $n_{\rm X}=2$ and 3 for the  
        solid (blue) and dotted (black) curve, respectively.}
      \label{fig:f8}
   \end{figure*}

For the {\it Suzaku} observation, the equivalent width (EW) of the
narrow 6.4 keV line is approximately 30 eV \citep{m07}, 
suggesting that the  distant reflector subtends a small solid angle at
the central X-ray source with $R=0.15$ being a likely value
characterising the strength  of the accompanying reflection hump.  In
Fig.\ \ref{fig:f7}a,  we show the RMS spectra for a simple
non-relativistic model  involving a variable power-law and a constant
reflection component with $R=0.15$ and a narrow Fe K$\alpha$ line with
EW=30 eV (both EW and $R$ are determined with respect to the average
power-law component).  The RMS spectra are computed for various levels
of variation of the power-law continuum; the index, $\Gamma$,  of the
power-law remains constant  while its intensity changes.  The
reflection spectrum is derived using the {\it pexrav} model, with the
inclination angle of the reflector fixed at $30 \degr$. The constant
line at 6.4 keV is computed with a rather large  width, $\sigma=0.1$
keV, for a clear  comparison with the observed RMS spectra (which are
derived with the energy bins $\Delta E = 0.2$ keV around 6.5 keV).
The constant line at 6.4 keV, with ${\rm EW}=30$ eV and $\sigma=0.1$
keV, is also included in all models in  Figs.\ \ref{fig:f7}(b),
\ref{fig:f8} and \ref{fig:f9}.

For a constant spectral component, with the fixed contribution to the
average spectrum, the strength of the related signature in the RMS
spectrum depends on the overall level of variability.  For ${\rm RMS}
= 0.3$, both a constant hump with $R=0.15$ and a line with ${\rm
  EW}=30$ eV would lead to rather pronounced  declines in the RMS
spectrum (see the top curve in Fig.\ \ref{fig:f7}a).  However, if
another spectral component dilutes the variability, as should be the
case in MCG--6-30-15, the strength of these declines decreases.
Specifically, ${\rm RMS}=0.05$--0.15, derived from the data above 10
keV, makes the hump with $R=0.15$ negligible. For ${\rm
  RMS}=0.15$--0.2, as observed  around 6 keV, the line with ${\rm
  EW}=30$ eV has a noticeable, but rather minor, effect. 

The observed shape of the RMS spectrum can be approximately
reproduced, above 3 keV,  in a simple non-relativistic model,
involving a constant reflection component with $R \ga 3$ (see
Fig.\ \ref{fig:f7}b).  Such an apparently unphysical (for reflection
from a distant matter) value of $R$ could  result from the effect of
the light travel time in a source observed in low luminosity state (as for
the {\it Suzaku} observation of NGC 4051; Terashima et al. 2009) or
from GR effects affecting the primary emission.  In the latter case,
more relevant for MCG--6-30-15, the mechanism is similar to model
$S^{\rm NZ}$.  Namely, if generation of X-rays occurs
within a few $R_{\rm g}$ from a rapidly  rotating black hole, a
face-on observer would directly receive reduced primary radiation,
which would be strongly focused  along the equatorial direction to the
distant material, giving rise to a strong non-relativistic reflection
component. In MCG--6-30-15  this scenario seems to be ruled out by the
small EW of the narrow K$\alpha$ line. Moreover, modeling the RMS
spectra for various time-bin sizes seems to require slightly
different values of $R$ (approximately 3 and 4 for 16 ksec and 131
ksec, respectively),  which is inconsistent with a simple reflection
model (where $R$ should not depend on the time-bin size).   As an
illustration, the RMS spectra in Fig.\ \ref{fig:f7}(b) are computed
with $R=4$.  While the lower curve roughly matches the  spectrum for
131 ksec, the upper curve strongly underpredicts the variability above
10 keV, as compared to the spectrum for 16 ksec.

\subsection{GR models}

\label{sect:4.3}

The RMS spectra predicted by the GR models for some ranges of parameters 
are qualitatively inconsistent with those observed, as discussed in Sect.\ 
\ref{sect:3}. Thus, the observed data completely rule out model $A$ (for completeness 
of our discussion, the best fit with this model is presented in Sect.\ \ref{sect:4.3.1}), 
model $S$ with $r_{\rm s}>4$ and model $C$ with $h_{\rm s}<2$, cf.\ Figs.\ \ref{fig:f1} 
and \ref{fig:f3}.

Some details of the RMS spectra in the GR models are sensitive to the
value of $\theta_{\rm obs}$. In our analysis 
we consider the inclination angles between $\theta_{\rm obs} = 25 \degr$ and  $45 \degr$.
We recall that the disc-line fits to the
{\it ASCA} observations of MCG--6-30-15, in which the line seems to extend
only up to $\simeq 6.8$ keV, yield $\theta_{\rm obs} \simeq 30 \degr$.
However, higher quality data indicate that the line extends beyond $7$
keV (but is indented by two absorption edges) and  recent fits to both
the {\it XMM} \citep{l07}
and {\it Suzaku} \citep{m07}
observations  indicate a larger value, $\theta_{\rm obs} \simeq 40 \degr$.

In models $A$ and $C$ we adjusted the amplitude of the RMS spectra to match the observed data, 
which is the usual 
procedure in spectral modeling. In model $S^{\rm NZ}$ we did not treat the amplitude as a free 
parameter. Instead, we attempted to reproduce self-consistently both the shape and the 
amplitude of the RMS spectrum by modifying the auxiliary parameters of the model, 
i.e., $\sigma_{\rm L}$ and $\delta$ or $\beta$.   

The model spectra best fitting the data  are shown in Fig.\ \ref{fig:f8}. 
Our formally best fitting case, shown by the solid curve in Fig.\ \ref{fig:f8}(b)
has the (reduced) $\chi_{\nu}^2 \approx 2$ (with 24 d.o.f.) for $E>3$ keV; 
the remaining fits in Fig.\ \ref{fig:f8} yield  $\chi_{\nu}^2 > 3$.
Clearly, the fits are not acceptable statistically.  
On the other hand, the basic trends are reproduced by model $S^{\rm NZ}$ reasonably well  
and we suspect that the quality of the fits could be improved 
when additional physical effects, neglected in this paper (e.g.\ ionization), are taken 
into account.

A further indication of the incompleteness
of our models is given by the systematic discrepancies below 4 keV.
Between 2 and 4 keV, the RMS spectra predicted by
the GR models are typically more convex than these observed, similarly
to the case of non-relativistic models (cf.\ Fig.\ \ref{fig:f7}b).
Only model $S^{\rm NZ}$ with $r_{\rm min}<2$ (producing a
strongly redshifted reflection component), yields the RMS spectra with
approximate agreement down to 2 keV.

Studies of strong gravity effects in the time-averaged spectrum of MCG--6-30-15 
usually concentrate on the data above 3 keV, as the spectrum at lower energies is strongly
affected by a warm absorber (while at larger energies the absorption
effects are considered to be unimportant; see e.g.\ \citealt{m07}).
We also do not attempt to model the RMS spectra below 3 keV, as changes of the absorber may
determine variability in this band. On the other hand, some
controversies remain as to whether the absorber  does exhibit strong
changes, see Sect.\ \ref{sect:5.6}, and we point out  that a weakly varying
absorption should not affect  the RMS variability significantly. Considering
the lack of absorber variability suggested by some previous
studies, we note that the deficiency of GR models below 3 keV
results from the small contribution to the total spectrum of the radiation
reflected from neutral matter at these energies. Then, ionization of the
innermost parts of the disc (which is most likely in model $S^{\rm NZ}$, 
see Sect.\ \ref{sect:5.5}) may be relevant to extend these models to lower energies. 

\subsubsection{Models $A$, $C$}

\label{sect:4.3.1}

In models $A$ and $C$ we considered $h_{\rm s} \le 20$. We were not
able to find the RMS spectra matching both the shape and the
normalization of those observed. Then,  we focused on reproducing the
shape of the RMS spectrum and we adjusted the amplitude, which is treated 
as a free parameter only in this section. 
 We presume that moderate changes of the RMS
normalization could result from a departure from the (vertically
constant)  distribution of the intrinsic luminosity assumed in these models.  
In general, the model RMS spectra with a (qualitatively) consistent shape 
have larger normalization than those observed, therefore, we neglected
variations of intrinsic luminosity in time, which would
lead  to a further increase of the RMS amplitude.  

The dashed curve in  Fig.\ \ref{fig:f8}(a) shows the RMS spectrum in model
$A$, with $\theta_{\rm obs}=35 \degr$, $h_{\rm min} = 2.2$ and $h_{\rm max} = 10$,
 best matching the observed shape; the model spectrum
is shifted down by $\Delta {\rm RMS} = - 0.4$. An
excess variability in the energy range of the Fe K$\alpha$  line, robustly 
predicted by model $A$, is clearly inconsistent with the observed data. 
We do not expect that modifications of our model, specifically those related 
to ionization of the disc, could reduce this discrepancy (see Sect.\ \ref{sect:5.5}) 
and thus we rule out this model. We also note that the observed level of
the RMS variability can be achieved in model $A$  with $h_{\rm min} > 3$, but in such a
case the model  RMS spectrum is much flatter than observed.
On the other hand, the stronger contribution from small $h_{\rm s}$, e.g.\ for $h_{\rm min} < 2.2$,
yields a larger excess around 6.4 keV.

Model $C$ predicts, for $\theta_{\rm obs} > 30 \degr$, a sharp drop at
the maximum of the blue peak, cf.\ Fig.\ 1(c), which cannot account for the observed,
broader depression in 6--7 keV range.  However, for $\theta_{\rm obs}=35
\degr$, the drop occurs around 6.8 keV and including an additional
narrow line at 6.4 keV we achieve an overall form of the depression
between 6 and 7 keV which can imitate the observed one. Then, we
found the set of parameters ($\theta_{\rm obs}=35 \degr$, $h_{\rm min} = 4$,
$h_{\rm max} = 20$; the solid curve in
Fig.\ \ref{fig:f8}a) for which deviations between the  predicted and
observed shapes  are small. The predicted RMS
amplitude only slightly exceeds the observed value, the model spectrum
is shifted down by $\Delta {\rm RMS} = - 0.07$. The dotted curve in
Fig.\ \ref{fig:f8}(a) illustrates the extrapolation of the
model for larger time-bin sizes, obtained with $n_X > 1$. 

Model $C$ requires a
fine-tuning of parameters to approximately explain the observed
fractional variability.  For the inclination angles higher or lower  than
$\theta_{\rm obs}=35 \degr$, discrepancies around the energy of the Fe
K$\alpha$ line are more significant. For the range of heights different
to $h_{\rm s} \simeq 4$--20,  the RMS spectrum
between 3 and 20 keV is too flat. See Sect.\ \ref{sect:5.3} for a further, 
critical discussion of the fits with model $C$.

\subsubsection{Model $S^{\rm NZ}$}

\label{sect:4.3.2}

Both the shape and amplitude of the observed RMS spectrum can be approximately
reproduced with $a \ge 0.98$ and the  distance of the primary source
varying in the range  extending down to at least $ r_{\rm s} \approx
2$. For smaller spins, or larger distances, the model RMS
spectrum is flatter  than observed.  Models with $n_X > 1$ strongly
favour the maximum value of $a=0.998$.  The
major challenge in modeling the observed RMS spectrum results from the
excess related to changes of the red wing,  which is not seen in the
data. A strong ionization of the disc, at the site of the formation of the variable
part of the red wing,  may be relevant in reducing this
discrepancy, see Sect.\ \ref{sect:5.4}. In the simplest case (with, in particular, a neutral and
untruncated disc) considered  in this paper, fits with model $S^{\rm
  PT}$ favour $r_{\rm min} < 2$, for which (1)
the excess is more extended and less pronounced, making a relatively
small deviation between the model and the data, and (2) the significant
contribution from the smallest $r_{\rm s}$ allows us to reproduce
the RMS spectra down to 2 keV. 

We attempted to reproduce the RMS spectra for both 16 and 131 ksec with 
the same set of parameters, differing only by the value of $n_{\rm X}$.
In our procedure, we first adjusted $r_{\rm min}$, $r_{\rm max}$, $n_{\rm X}$,
$\delta$, $\sigma_{\rm L}$ and $\theta_{\rm obs}$ to fit the 16 ksec spectrum, 
then we computed the spectra for larger $n_{\rm X}$ and compared them to 131 ksec. 
The differences between the amplitudes of the RMS for consecutive $n_{\rm X}$ are rather 
large, which is most challenging in attempts to reproduce both time-bins simultaneously
(as noted before, the underlying assumption of identical $n_{\rm X}$ for each time bin 
is most likely oversimplified).

Our best fits, with $a=0.998$, $\theta_{\rm obs}=35 \degr$, $r_{\rm min}=1.6$,
 $r_{\rm max}=3$, $\sigma_{\rm L}=0.2L^{\rm PT}$, $\delta=-1.5$, $n_{\rm X}=3$ and 5 
for 16 and 131 ksec, respectively, are shown in Fig.\ \ref{fig:f8}(b).
Disregarding the excess variability of the red wing  (with the plausible impact of
ionization in mind) we can reproduce the observed  RMS spectra for a broader
range of parameters. Example fits, achieved using $\beta$ instead of $\sigma_{\rm L}$ and
$\delta$, with $\theta_{\rm obs}=35 \degr$, $r_{\rm min}=1.8$ 
and $r_{\rm max}=3$, $\beta=1$, $n_{\rm X}=2$ (for 16 ksec) and 3 (for 131 ksec) 
are shown in Fig.\ \ref{fig:f8}(c).

The statistical quality of the fits is poor, as noted above, therefore,
the usual criterion for constraining the confidence limits in the model parameters,
using the increase of $\chi^2$, seems inadequate. Qualitatively, 
$r_{\rm min} \simeq 2.2$ seems to be the upper limit for $r_{\rm min}$; 
for larger $r_{\rm min}$,  the RMS spectrum is too flat, see the bottom 
curve in Fig.\ 5(c). Similarly, we can reject submaximal ($<0.98$)
values of $a$, cf.\ Fig.\ 4(a). The model with
$r_{\rm min}=1.6$ and  $r_{\rm max}=3$ is slightly favored in our fits;
 Fig.\ \ref{fig:f8}(b) shows trends corresponding to the extention of this range. 
In general, models with $r_{\rm max}>3$ give  a more significant excess with respect 
to the data below 5 keV, see the dot-dashed curve.
In turn, for our best fit with $r_{\rm min}=1.4$, 
variability above 10 keV is slightly underpredicted; notably, this fit (the dashed curve) 
involves no modification of our basic $S^{\rm NZ}$ model, i.e.\ $\sigma_{\rm L}=0$, 
$\delta=0$ and $\beta=0$, so the intrinsic luminosity follows exactly $L^{\rm PT}$.
See Sect.\ \ref{sect:5.4} for further discussion of $r_{\rm min}$ and $r_{\rm max}$.

   \begin{figure}
   \centering \includegraphics[width=7cm]{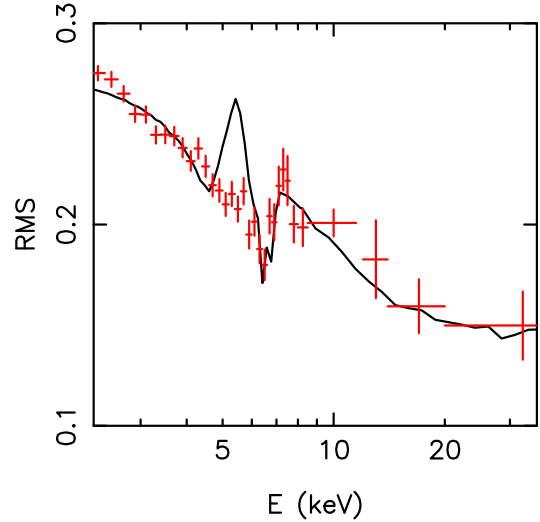}
      \caption{RMS spectrum for model $S^{\rm NZ}$ with an excess
        around 5.5 keV,  as revealed by some previous observations
        (see text).  Parameters  $\beta=1.2$, $\delta=0$, $\sigma_{\rm L}=0$
        and $n_X=1$ are adjusted to match the {\it Suzaku}
        RMS spectrum (for 16 ksec) at $E<5$ keV and $E > 6$ keV.  }
         \label{fig:f9}
   \end{figure}

\subsubsection{Excess variability between 5 and 6 keV}

\label{sect:4.3.3}

As we discuss in Sect.\ \ref{sect:5.5}, the excess variability of the red wing,
which is the major prediction of model $S^{\rm NZ}$ ruled out by the {\it
  Suzaku}  data, is likely to be  reduced by
ionization of the disc  under the source.  Interestingly, however, the
RMS spectra from some previous observations of MCG--6-30-15 \citep[see][]{mifi03,p04}
did reveal a strong
enhancement of the RMS variability between 5 and 6 keV. Apart from
this band, those RMS spectra are consistent with other observations of
MCG--6-30-15.  Presumably, the geometry of the X-ray emitting region
does not change significantly between the various  observations  and the
presence or lack of an excess is related to changing ionization
state of the innermost disc.

Then, we examined conditions in which model $S^{\rm NZ}$ produces an
excess between 5 and 6 keV but matches the {\it Suzaku} RMS spectra at
other energies.  An example of such an RMS spectrum for $\theta_{\rm
  obs}=40 \degr$, $r_{\rm min} = 2.2$ and $r_{\rm max} =3.2$  is shown in
Fig.\  \ref{fig:f9}.  We find that, in general, a relatively  high
inclination angle, $\simeq 40 \degr - 45 \degr$, is required to explain   the
excess between 5 and 6 keV by a varying redshift of 6.4 keV photons --
for smaller $\theta_{\rm obs}$ the excess occurs at lower energies. 

The amplitude of the excess in Fig.\ \ref{fig:f9} is approximately
consistent with that  found by \citet{p04}
in the {\it XMM} data. However, we were not able to reproduce a stronger excess, with
${\rm RMS} \approx 0.4$,  as reported by \citet{mifi03}
for the {\it ASCA} observation. Such a strong excess can be explained by
an enhanced contribution of Fe photons from the spot  under the source,
which could occur if ionization of that area of the disc
is taken into account, namely, it
could be attributed to  intermediate stages of ionization  (where 
Auger destruction cannot operate) giving rise to an intensity of the Fe
photons  much higher than from a neutral medium \citep[e.g.][]{zcz94}.  
For these ionization states, the rest energy of Fe
photons is around 6.9 keV, and the range of inclination angle  
relevant to reproduce the excess would be
$\theta_{\rm obs} \simeq 35 \degr - 40 \degr$. 

Interestingly, during the {\it XMM} observation analysed  by Ponti et al.\ (2004),
the source luminosity was
$\sim 2$ times lower than the typical luminosity observed in MCG--6-30-15.
This can support the above scenario, with the change from complete to intermediate
ionization of a hot spot.  In case of two {\it ASCA} observations used by Matsumoto 
et al.\ (2003) to reveal enhanced variability between 5 and 6 keV, the average
luminosity was not much lower than typical, however, both observations contain
prolonged deep minimum states, with the luminosity dropping to the lowest level 
observed in that object.

We point out that an excess between 5 and 6 keV cannot be produced by
simple geometric  effects in models $A$ or $C$; actually,  the latter predicts
an opposite property, i.e. reduced variability around 5.5 keV.

\section{Summary and discussion}

\subsection{Conclusions}

This paper focused on modeling the RMS spectra, which are commonly
used to estimate the fractional variability amplitude in AGNs.  We
thoroughly studied the RMS spectra predicted by models relating
spectral changes to varying amounts of GR effects, which distort the
X-ray radiation generated in central parts of black-hole accretion flow. We
applied our GR models to MCG--6-30-15 where the bulk of the
reprocessed radiation is supposed to come from the inner accretion disc.  Our main
conclusions are summarized as follows:

\noindent
(1) The Compton reflection hump, reprocessed  in the accretion disc
from the continuum emission generated at a small, varying
distance from a black hole, is typically less variable than the
primary continuum. This leads to a decrease of the RMS spectrum
with increasing energy, above 3 keV (for neutral disc
material).  A remarkable exception involves a source corotating with
the  disc and changing height (as in the model of MF04) not far above the disc surface 
-- in this case the fractional variability increases with energy.

\noindent
(2) Relativistic distortions of the Fe K$\alpha$ line give rise to
excess variability  which is the major challenge in applications of
GR models to observed data. In models involving vertical motion of
the X-ray source, the excess variability, between 6 and 7 keV, is related to changes of the
blue peak. In models involving changes
of the radial distance, low above the disc surface,  the excess is related
to changes of the red wing and occurs between 4 and 6 keV. In the
latter class of models, however, ionization of the hot spot under 
the source is very likely and should reduce the
excess variability, see Sect.\ \ref{sect:5.5}.

\noindent
(3) The results of this paper illustrate the maximum strength of
signals in the RMS spectra caused by the GR effects. Such strong signals could be
observationally seen only in the RMS spectra determined with time-bin sizes
not exceeding the time scale of the change of position of the X-ray
source. For longer time bins,
signatures of GR effects are weaker (i.e., the RMS spectra are flatter);
some implications of this are discussed in Sect.\ \ref{sect:5.2}.

\noindent
(4)  Variability observed in MCG--6-30-15 is inconsistent
with models assuming vertical motion of the source. The model
with a source on the symmetry axis is ruled out and the
model by \citetalias{mf04} involving a specific pattern of
rotation around the symmetry axis is disfavored as discussed in Sect.\ \ref{sect:5.3}. 

\noindent
(5) The model with changes of the radial distance in
the Kerr metric, proposed in \citetalias{nz08}, offers the most likely explanation of the 
observed properties (among the GR models); See Sect.\ \ref{sect:5.4} for further
discussion of this model.

\noindent
(6) Some details of the RMS spectra are sensitive to $\theta_{\rm
  obs}$, and they may provide constraints on inclination, independent
of the modeling of the line profile.  Interestingly, our GR fits to
the RMS spectra from {\it Suzaku} observations of MCG--6-30-15 favor
$\theta_{\rm obs}=35 \degr - 40 \degr $, consistent with  the disc-line fits
to the {\it Suzaku} data.  The light-bending \citepalias{mf04}
model requires exactly $\theta_{\rm obs}=35 \degr$.  The best fit with model 
$S^{\rm NZ}$ also has $\theta_{\rm obs}=35 \degr$, however, this model  
allows for a broader range of $\theta_{\rm obs}$, especially
when excess due to  variability of the red wing is disregarded.
The range of $\theta_{\rm obs}=35 \degr -40 \degr $ would also be relevant in
reproducing the excess variability between 5 and 6 keV, which was
reported for some  previous {\it ASCA} and {\it XMM} observations, in
a model involving a highly ionized hot spot under the hard X-ray source.

\subsection{Variability time-scales}
\label{sect:5.2}

The energy spectra determined for increasing lengths of time-bins
should consist of contributions 
from an increasing  number of locations of the primary source. 
Then, the RMS spectra for longer time bins should be flatter, see
Sect.\ \ref{sect:3.4}. Below we briefly discuss the resulting implications   for
time-scales of various processes involved in generating the observed variability.

The reduced variations of the reprocessed component are typically
assessed over $\ga 10$ ksec. We emphasize that the actual
time-scale of the change of the position of the hard X-ray source should be
at least of this order of magnitude, otherwise the RMS spectra would be 
much flatter than observed. Then, significant changes of the X-ray flux on a
time-scale of $\sim 1$ ksec which are observed in MCG--6-30-15
should result from intrinsic variations
(with small changes of geometry) of the hard X-ray source.
The point-to-point RMS spectrum with 1 ksec time
bins which measures such short time-scale variability, found to be much
flatter than the standard RMS spectra in \citet{vf04},
supports the scenario with changes of intrinsic luminosity
dominating the short time-scale  variability (by definition of the RMS spectrum, 
changes  of luminosity, without the change  of spectral shape, produce a flat RMS spectrum).

In Sect.\ \ref{sect:4.3.2} we note that the RMS spectra for both 16 ksec
and 131 ksec can be reproduced with the same set of parameters, in a model 
involving a larger number of the source positions for
the larger bin size.  However, the number of positions in our fits for 131 ksec
is only $\sim 2$ times larger than for 16 ksec, while  a
factor of 8 would be expected for a simple scaling of the number of positions
with the bin length. Such a simple, linear scaling of the number 
of source locations is ruled out, as $\ga 10$ random locations, implied
for 131 ksec, would yield an almost flat RMS spectrum.  
Then, the change of geometry on a time scale of $\sim 100$ ksec should
obey  some systematic pattern, involving at most a few localised
regions dominating the total emission, with the duration of several tens of ksec.
Emission averaged over the whole
range of locations of the source is probably probed on a much longer
time-scale, $> 1000$ ksec, e.g. by the RMS spectra derived from
long-term {\it RXTE} monitoring, see \citet{m03},
which are indeed flatter than these obtained with $10-100$ ksec bins.

Finally, note that 
the time-scales for the change of the source location, $\sim 10-100$ ksec, required by the 
above arguments are orders of
magnitude longer than the dynamical time-scale for the innermost
region, $\la 100$ sec, which is rather puzzling. On the other hand,
the power-spectra indicate that most of the variability  in
MCG--6-30-15 indeed occurs on time scales $\ge 10$ ksec and  a
large-magnitude  variability may extend even to time-scales longer than $100$
ksec \citep[see][]{vfn03,u02}.
Then, in the context
of GR models, most of the power contained  in the X-ray light curve of
MCG--6-30-15 would result from changes of the geometry of the
X-ray source. In turn, variability 
dominated by changes of intrinsic luminosity should correspond to
frequencies higher than the break frequency ($10^{-4}$ Hz, above which
the magnitude of variability drops).

\subsection{Models involving vertical motion}

\label{sect:5.3}

Focusing of radiation produced 
close to the symmetry axis, toward the disc, cannot explain the detailed properties 
of MCG--6-30-15, despite qualitative arguments for the relevance of this mechanism. 
Significant changes of the blue peak corresponding to changes of the source height 
are the major cause of this model failure.

This discrepancy can be reduced in a modification of the model by MF04, 
where the observed RMS spectrum can be approximately reproduced with the height 
changing between $\simeq 4$ and $20 R_{\rm g}$.
We emphasize again that predictions of this light-bending model are related primarily
to its specific kinematic  assumptions rather than the light-bending
itself. We disfavor this model on the grounds  that it
requires large elevations of the source above the disc, where
the kinematic assumptions are rather arbitrary. At lower
elevations (where the crucial assumption of this model -- i.e.\
rotation of the source with $\Omega_{\rm K}(\rho_{\rm s})$, see Sect.\ \ref{sect:2.1} -- 
has some physical motivation), the model predicts a variability pattern that is ruled out
by observations.
Even with the required range of $h_{\rm s}$, the model is rather unlikely
as a decline of the fractional variability  around  5.5 keV, systematically
predicted by the model, has not been confirmed by any
observation of MCG--6-30-15 (actually an opposite  property has been
reported, see Sect.\ \ref{sect:4.3.3}). Another major shortcoming 
of this model was pointed out by \citetalias{nz08}. For
the range of heights relevant to reproduce the observed variability ($\ge 4 R_{\rm g}$), 
the model cannot explain the pronounced red wing of the relativistic
Fe line, the feature that originally motivated the development
of this model.

\subsection{Model with bending to the equatorial plane}

\label{sect:5.4}

We strongly  favor the model with radial motion in the Kerr metric for a description
of the central region of MCG--6-30-15.
Note that both of the main properties required by this model, i.e.\ rapid rotation 
of the black hole and very small distance of the X-ray source, have some independent 
support in the observed, time-averaged Fe line profile. 
Our condition for the value of the spin, $a>0.98$,
is consistent with the formal constraint on the spin
parameter, $a > 0.987$ \citep{br06}, derived from the
line profile  under the assumption that the Fe photons are emitted only
at distances larger than $r_{\rm ms}(a)$.
Regarding the second condition (small distance), \citetalias{nz08} point out that 
location of the X-ray source within a few central  $R_{\rm g}$
is needed to explain  the pronounced red wing of the 
line profile; the irradiation of the inner disc by a source located at further distances 
is too weak and the red wing is much weaker relative to the blue peak than in the observed profile.

In the relevant range of distances, the
directly observed flux of primary X-rays is extremely sensitive to the location
of the X-ray source, while the reflected radiation changes a little, and
the observed variability effects can be reproduced by rather small changes in
the geometry of the X-ray emitting region. The range of distances between approximately 
$1.6 R_{\rm g}$ and $3 R_{\rm g}$ is slightly preferred in our applications of this
model to the {\it Suzaku} data.
However, we emphasize again that some features of our model are oversimplified,
most importantly our method for the generation of 
active regions and their luminosities as well as our neglect of ionization,
and more realistic modeling could yield a more extended range of locations. 

Remarkably, for $a=0.998$, the maximum of the dissipation rate given by the \citet{pt74}   
formula occurs exactly at the lower end (i.e.\ $1.6 R_{\rm g}$) of this range. 
Furthermore, according to that formula with $a=0.998$, $35$ per cent of the total power dissipated
in a Keplerian disc is released within the central $3 R_{\rm g}$. Then, our condition for 
the location of the hard X-ray source, although seemingly rather extreme, is not implausible.

An independent  constraint on the range of distances 
comes from the normalization of the reflection component in the time-averaged spectrum. 
We have computed the reflection parameter, $R$, in our models as the ratio of the total fluxes 
in the reflected component in the GR model and in the {\it pexrav} model with the same flux of the primary power-law. Our fits with model $S^{\rm NZ}$, shown in Fig.\ \ref{fig:f8}, have $R \ga 4$,
within the confidence limits, $R=3.8 \pm 0.7$, found for {\it Suzaku}
observations in \citet{m07}.
A stronger contribution of primary hard X-rays from smaller radii (e.g.\ for  
$r_{\rm min} < 1.6$, $r_{\rm max}=3$ and $\delta < 0$)
would yield, however, a larger value of $R$, inconsistent with the data.

In our investigation of the model we focused on the case with $h_{\rm s}/r_{\rm s} < 0.1$. 
For $h_{\rm s}/R_{\rm s} \gg 1$, i.e.\ in the lamp-post regime, the bending to the center dominates
over the bending to the equatorial plane, leading to qualitatively different effects, as 
described above. We did not investigate the case of intermediate $h_{\rm s}/r_{\rm s}$, 
as results in this regime strongly depend on (very uncertain) kinematic assumptions. 
Note that even for relatively small $h_{\rm s}/r_{\rm s} = 0.2$, the difference between 
the azimuthal velocities resulting from the different assumptions  
discussed in Sect.\ \ref{sect:2.1} exceeds $0.1c$.

Note that occasionally an additional source of primary X-ray
emission may be present in MCG--6-30-15, which would lead to a more complex 
temporal behaviour than estimated in this paper. Flaring releases of a part 
of the X-ray emission may occur at rather large heights ($> 100 R_{\rm g}$, 
note that GR effects are negligible at these distances). We note the following 
hints for such episodic releases. MCG--6-30-15 typically
shows no evidence that variations in the Fe line track the continuum
variations, which supports the scenario with X-rays released and
reprocessed in the very central region.  However, during very  strong flares (in the light-curve),  
the line appears to respond to changes of the X-ray
flux and lags of a few ksec are observed between the flare and the appearance of 
an enhanced Fe line \citep{p04,n}.
A likely explanation is given by the strong X-ray flare
at a height of several hundred $R_{\rm g}$. In agreement with such scenario,  
the (delayed) Fe line became narrower in time as if it was coming from more 
distant regions of the disc, in a manner consistent with a response to 
the flare that occurred at a height of several hundred $R_{\rm g}$.  

Finally, we note a caveat against the specific scenario
considered in this paper, with hard X-rays produced by (at most)
several compact sources corotating with  the disc. A
quasi-periodic modulation of the primary emission should appear in
such a case on the Keplerian time scale, $\sim 100$ sec, however,
such a signal is  not observed in MCG--6-30-15 (see \citealt{zn05}
for details). This
may imply that the hard X-ray source has a more continuous spacial distribution,
e.g.\ forms an extended corona covering a part of the disc surface or a small hot
torus replacing  the innermost disc. If such an extended, hot plasma is 
located in a very small central region and exhibits small changes of geometry, 
the mechanism described in Sect.\ \ref{sect:3.3} should lead to properties 
qualitatively consistent with these derived for model $S^{\rm NZ}$ -- see \citet{nf07}
for preliminary results of the model with a small, shrinking/expanding inner torus. 

\subsection{Ionization}
\label{sect:5.5}

The neglect of the ionization of the disc surface is the major shortcoming of our model. 
Determination  of the ionization structure requires rather complicated radiative transfer 
computations \citep[see, e.g.,][]{r02}
which are currently not included in our model. 
For MCG--6-30-15, it seems to be well established that the blue peak
of the Fe K$\alpha$ line arises from neutral Fe at a distance of
$\ga 10R_{\rm g}$ \citep[e.g.][]{vf04}.
However, the inner parts of the disc, where the red wing
is formed, may be highly ionized \citep[cf.][]{b03}. 
The irradiation pattern in model $S^{\rm NZ}$ should yield
a non-uniform ionization structure of the disc, with a strongly  
ionized inner region and neutral outer
parts, as for the relevant range
of distances, $r_{\rm s}=2-3$, the flux irradiating the spot
immediately under the source is $\sim 4$ orders of magnitude
larger than that  irradiating the disc at $\sim 10 R_{\rm g}$. 

A precise estimation of the impact of ionization effects on the Fe line 
variability is difficult without detailed computations, as
a compact X-ray source above the inner disc should produce 
a range of ionization states over the surface
of the disc and the profile of ionization
should change in response to the change of location of the source.
Qualitatively, however, we may expect that fits of the observed RMS 
spectra would be improved with these effects taken into account, 
in particular the major discrepancy between model $S^{\rm NZ}$
and the {\it Suzaku} data, i.e.\ the excess variability in the red wing, 
could be reduced. This would require an ionization state  
of the hot spot under the hard X-ray source (where the variable component of the
line originates when the disc is neutral) corresponding 
either to resonant trapping of Fe photons or to complete ionization of
the Fe atoms (see, e.g., \citealt{zcz94} for details).

A remarkable consequence of ionization, crucial for modeling MCG--6-30-15,
can be expected also in the soft X-ray band. 
Reflection from ionized material is more efficient at low energies
than reflection from neutral matter, which may be relevant
in modeling the low energy part of the RMS spectra, typically not
well  described below 3 keV by neutral reflection models. Alternatively, changes of 
a warm absorber could explain variability in this band, see next section;
distinction between these two models with ionized reflection or warm
absorption is somewhat ambiguous (in soft X-rays),
as the photoionized disc reflection model  reproduces many features in the
spectrum that could otherwise be interpreted as warm absorption
edges \citep{c06}.

Our speculative picture for model $S^{\rm NZ}$ (with an ionized innermost and neutral 
outer disc) is similar to the model involving two
distinct reflection regions  on the disc, a highly ionized inner
region at a few $R_{\rm g}$ and a neutral outer region at larger
distances, applied to {\it XMM} observations of MCG--6-30-15 by \citet{b03}.
Interestingly, such a model can also describe the X-ray
spectra  of other high-accretion rate AGNs \citep[see e.g.][]{s07}.

The discrepancies predicted by models with vertical motion result mostly from 
the change of the blue peak, which should come from a neutral disc (see above). 
Then, ionization effects appear to be not relevant in improving their applicability.
Furthermore, in these models irradiation of the disc is much more uniform both 
azimuthally and radially, e.g.\ for $h_{\rm s} \simeq 10$  the flux
irradiating the disc around $6 R_{\rm g}$ is only larger by a factor of 2--3
 than that around $10-20 R_{\rm g}$, and, therefore, a range of ionization 
states along the disc surface, which could give rise to 
more complex variability scenarios, is unlikely. Finally,  
ionization in the inner part, which  could be considered to solve the problem 
of the RMS decline around 5.5 keV predicted by the \citetalias{mf04} model seems to 
be ruled out (again due to the very uniform illumination) as ionization in that part 
would lead to strong reduction of the whole red wing (which in this model is
too weak even for a neutral disc, see above),
contrary to model $S^{\rm NZ}$, where the azimuthal distribution of the irradiation 
of the inner disc is very non-uniform and a
strong red wing can be explained even if its variable part is
depleted by the strong ionization of a hot spot.

\subsection{Alternative (non GR) models of MCG--6-30-15}

\label{sect:5.6}

The development of GR variability models was motivated by
detections of strongly relativistic  distortions of the Fe line. A
further motivation is the fact that the model including 
a strong distant reflection, which could explain some of trends
observed in MCG--6-30-15, is not applicable to this object
(Sect.\ \ref{sect:4.2}).

The remaining alternative involves  a variable absorption which can
produce  an energy dependent RMS variability, with larger variations
occurring at these energies where the flux is affected by stronger
absorption.  The basic picture was developed by \citet{i03},
who postulate changes of the column  density of an absorber in
MCG--6-30-15 on a time scale of $\sim 10^4$--$10^5$ sec,  under the
assumption that the absorption cross section does not change.  In this
scenario, reduced variability between 5 and 7 keV is explained by
higher transparency  of the absorber in this energy range. Similarly,
an excess in the spectrum in this band -  commonly interpreted as a
disc line - is attributed to the small opacity of the absorber.  Note,
however, that the stronger decrease of the RMS variability around 6
keV for larger time bin widths,  which motivated the model proposed by
Inoue \& Matsumoto, was seen only in one {\it ASCA}
observation.

Several contradictions to  such a model were pointed out based on the
{\it XMM} observations  (see \citealt{vf04}
for a detailed discussion, disfavouring interpretations of the spectral shape
excluding the presence of the disc line and ruling out explanations of
the variability in terms of complex absorption). Most importantly, an
analysis of the absorption lines for the 2001 {\it XMM}  observation
did not reveal any substantial changes of the absorber \citep{t04}
which appears to be constant while the spectrum exhibits
changes typical for this source.

Recently, a revised model involving multiple absorption zones has
been proposed by \citet{m08}
and it seems that the contribution to the variability coming from 
changes of the  absorber remains an uncertain issue. 
Obviously, photoelectric absorption is unimportant above 10 keV and
other reduction mechanisms should be taken into
account if the reduction of variability extends to several tens of
keV, as  concluded by \citet{m07}.
In terms of non-GR models, such a reduction could be explained if changes 
of the spectral index were present
with  a pivot at $E > 10$ keV. Fits to the time
resolved spectra with the two-component model do not reveal  such trends 
(except for a few cases with the lowest flux, see e.g.\ \citealt{l07}); 
however, they assume the presence of the strong reflection component
and, therefore, this conclusion may be model-dependent. 
A model involving  short-term variations of 
the ionization degree  of warm absorbers, combined with changes of the spectral index,
for the {\it Suzaku} observations is currently under investigation 
(Miyakawa \& Ebisawa, in progress). 

\subsection{Other objects}

AGNs show a range of spectral variability properties. Some show trends
similar to  MCG--6-30-15 but these are certainly not ubiquitous.
Accretion rates and black hole masses, similar to those in MCG--6-30-15,
are typical for Narrow Line Seyfert 1 (NLSy1) galaxies. 
A similar geometry of the central region may be common in these objects.
Indeed, NLSy1s often show reflection-dominated spectra. Moreover,
the blurred reflection  fits require that  emission from within a few
$R_{\rm g}$ dominates, which makes them  similar to MCG--6-30-15;
however, in some of them the RMS spectra are flat \citep[e.g.][]{z08}.
On the grounds of our results (see Fig.\ \ref{fig:f4}a),
we point out that such a difference may result from slightly different
speeds of black hole rotation, with the maximal ($a > 0.98$) speed
tentatively expected in  MCG--6-30-15, and submaximal ($a \la 0.95$) which
may characterize other AGNs. An object with the same geometry of the
innermost region as favoured here for MCG--6-30-15, but harbouring a
slightly submaximally rotating black hole, would produce a
reflection-dominated spectrum, subject to extreme relativistic
blurring (as $r_{\rm ms} \approx 2$), but its RMS spectra would be flat.

Recent studies of the X-ray spectra in AGNs involving  the
blurred reflection model often indicated rapid rotation  of
supermassive black holes; moreover, the radial emissivity  index of
the fits indicates that emission from  the innermost (within a few $R_{\rm g}$) 
part of the disc dominates \citep[see e.g.][]{c06}.
Note that the relativistic blurring
in such models is typically based on a \citet{l91} line profile, which
assumes a maximal value of $a=0.998$, so the value of $a$ is not
measured and the fits requiring an extreme rotation of the black hole
do not exclude a slightly submaximal value of $a$ (which, following the above
arguments, may determine the observational differences between MCG--6-30-15 
and other objects from this class).

NGC 4051 is an example of NLSy1 with a spectral variability similar to
that in MCG--6-30-15, at least in the higher flux states \citep{p06}.
The two-component model with the constant reflection component involving 
the contribution from an ionized inner disc used by \citet{p06}
to explain variability at higher fluxes is
analogous to that describing MCG--6-30-15. Interestingly, in the low flux
state studied by \citet{p06}
the variability pattern is different and -- in the context of GR models -- 
it may imply that the
low flux states are related to prolonged periods with the hard X-ray
source located in the position closest to the black hole and
exhibiting moderate intrinsic changes. However, as discussed in \citet{t09},
these low flux states may be strongly
dominated by spectral components formed in distant material, which
makes investigation of the innermost region less straightforward.

The RMS spectra decreasing at higher energies, with depressions around 6 keV, 
are observed in many AGNs \citep[see e.g.][]{m03} 
and the two-component model, involving the reflection component  constant both
in normalization and in spectral shape, has been popular recently in the
modeling of such spectra. The first of the observed properties, i.e.\ the decrease with 
increasing energy, is predicted by most GR models, however, it
requires a rather small distance, $\la 10 R_{\rm g}$, of the X-ray source, 
which is not a likely approximation in some cases, e.g.\ for those objects 
where an extended ($\sim$ a few tens of $R_{\rm g}$), hot flow is supposed to form. 
Moreover, all GR models predict some changes in shape 
and, therefore, results achieved with the two-component model
cannot be directly interpreted in terms of GR effects.
In turn, the second of the observed properties, i.e.\ depression around 6 keV, 
is not easily reproduced by GR models and its explanation 
requires even more extreme parameters, namely rapid rotation (with $a > 0.95$, 
cf.\ Fig.\ 4a) and a distance $\la 3 R_{\rm g}$.

A strictly constant component can be produced by reflection from distant
matter; however,  a strong contribution of such radiation is not supported
by recent models (\citealt{m09};
note that our Fig.\ \ref{fig:f7}a illustrates the strength of effects 
corresponding to the amount of reflection from a distant torus 
estimated in that paper,
i.e.\ 6 times weaker than from the {\it pexrav} model with $R=1$). 
Furthermore, such a strong contribution would yield
pronounced, sharp drops at 7 keV  both in the RMS spectrum (see
Fig.\ \ref{fig:f7}b) and in the average energy spectrum.  The latter
is only occasionally observed in AGNs (e.g.\ \citealt{b02};
interestingly, the RMS for this observation is flat). 

\begin{acknowledgements}
We are grateful to the referee for numerous comments which 
helped us in improving this paper. This work was partly supported
by a grant N203 011 32/1518 from the Polish Ministry of Science 
and Higher Education.

\end{acknowledgements}

\bibliographystyle{aa}

\label{lastpage}
\end{document}